\documentclass[reprint,aps,prb,amsmath,amssymb,superscriptaddress]{revtex4-2}

\usepackage{graphicx}
\usepackage{bm}
\usepackage{lipsum}   
\usepackage[hidelinks]{hyperref} 
\usepackage{physics}

\begin{document}
\title{Exciton Polariton–Polariton Interactions in Transition-Metal Dichalcogenides}
\author{Jonas K. K\"onig}
\email{jonas.koenig@physik.uni-marburg.de}
\affiliation{Department of Physics, Philipps-Universit{\"a}t Marburg, 
35037 Marburg, Germany}
\affiliation{mar.quest|Marburg
Center for Quantum Materials and Sustainable Technologies,
35032 Marburg, Germany}
\author{Jamie M. Fitzgerald}
\affiliation{Department of Physics, Philipps-Universit{\"a}t Marburg, 
35037 Marburg, Germany}
\affiliation{mar.quest|Marburg
Center for Quantum Materials and Sustainable Technologies,
35032 Marburg, Germany}
\author{Daniel Erkensten}
\affiliation{Department of Physics, Philipps-Universit{\"a}t Marburg, 
35037 Marburg, Germany}
\affiliation{mar.quest|Marburg
Center for Quantum Materials and Sustainable Technologies,
35032 Marburg, Germany}
\author{Ermin Malic}
\affiliation{Department of Physics, Philipps-Universit{\"a}t Marburg, 
35037 Marburg, Germany}
\affiliation{mar.quest|Marburg
Center for Quantum Materials and Sustainable Technologies,
35032 Marburg, Germany}

\begin{abstract}
Microscopic insights into nonlinear interactions are essential for advancing polaritonic devices. Existing studies often rely on phenomenological models that overlook important many-body processes. Based on a material-specific and predictive approach, we investigate monolayer and homobilayer MoS$_2$ embedded in a Fabry-Pérot cavity to characterize the exchange, saturation, and dipole-dipole contributions to polariton-polariton interactions in these technologically promising materials. A key finding is that the exchange interaction induces asymmetric energy shifts of the lower and upper polariton branches in a detuned cavity, a behavior driven by the difference in their excitonic character. Furthermore, we demonstrate that temperature and electron–photon coupling determine the energy renormalization through the equilibrium polariton distribution. In homobilayers, the dipole-dipole interaction is mediated by the interlayer character, enabling electrical control and facilitating the electric-field-induced closing of anti-crossings due to dipolar-interaction shifts. The gained insights on polariton–polariton interactions are important for the development of ultra-compact polaritonic circuitry.
\end{abstract}
\maketitle

\section*{Introduction}
\begin{figure}[!t]
    \centering
    \includegraphics[width=0.5\textwidth]{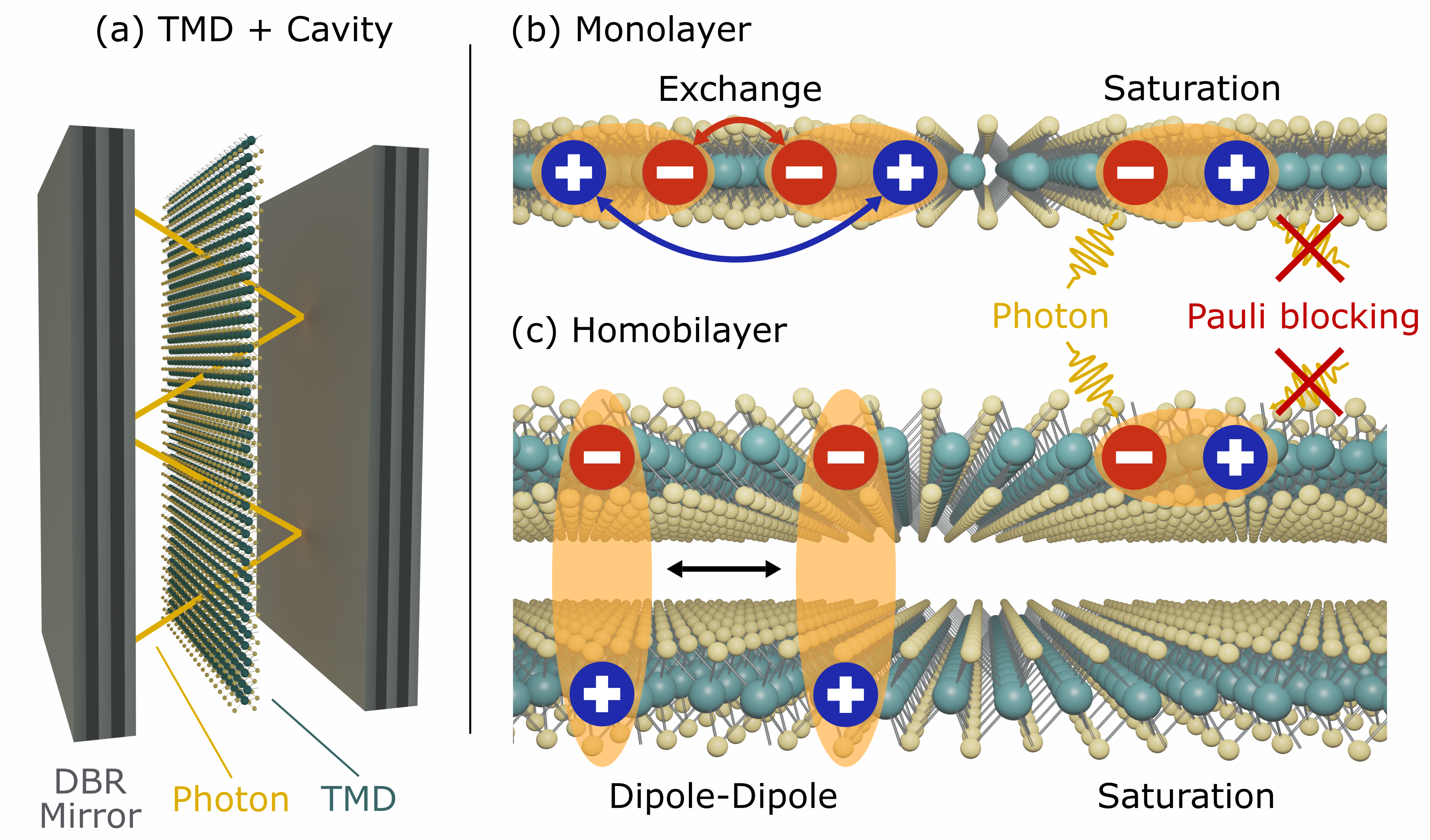}
    \caption{(a) Schematic of a TMD monolayer in a Fabry-Pérot microcavity and (b) polariton-polariton interactions derived from the constituent exciton interactions: (1) Exchange interactions between identical fermions and (2) saturation due to the Pauli exclusion principle. (c) The TMD homobilayer case with: (1) dipole-dipole repulsion between interlayer excitons and (2) intralayer exciton saturation.}
    \label{fig:1}
\end{figure}

Two-dimensional (2D) transition metal dichalcogenides (TMDs) have emerged as a versatile platform in materials science and optoelectronics, combining strong light–matter interactions \cite{dufferwiel2015exciton} with a rich exciton landscape \cite{mueller2018exciton,hagel2021exciton} for exploring intriguing correlations and many-body physics \cite{tempelaar2019many,merkl2020twist,trovatello2022disentangling}. These atomically thin semiconductors host excitons, Coulomb-bound electron-hole pairs with exceptionally large binding energies of hundreds of meV that persist even at room temperature \cite{chernikov2014exciton,palummo2015exciton,perea2022exciton}. When these TMDs are embedded within a microcavity (Fig.~\ref{fig:1}(a)), their robust excitons couple strongly with the confined photonic mode, forming quasi-particles known as exciton-polaritons \cite{schneider2018two,louca2023interspecies}. These hybrid light-matter states inherit the low effective mass and high mobility of cavity photons \cite{topfer2020time,fitzgerald2025polariton}, while retaining the strong, inherent nonlinear interactions and field tunability of excitons \cite{dufferwiel2018valley,zhang2021van,datta2022highly,louca2023interspecies,nakano2024light}. These nonlinearities enable ultrafast control of exciton-photon coupling, including femtosecond switching of the Rabi splitting \cite{genco2025femtosecond}, underscoring the potential of polaritons for future integrated low-threshold lasers, polaritonic circuits, and quantum information processing \cite{kavokin2022polariton}. Polariton nonlinearities have been studied for decades in quantum well systems \cite{rochat2000excitonic,glazov2009polariton, rhee1996femtosecond} and have more recently been investigated in other high-performance materials, such as organic semiconductors \cite{daskalakis2014nonlinear,wang2021large}, lead halide perovskites \cite{fieramosca2019two,su2020observation,masharin2022polaron}, and TMDs \cite{kravtsov2020nonlinear,gu2021enhanced,zhang2021van,datta2022highly,louca2023interspecies,xiang2026electrically}
 including moiré systems \cite{zhang2021van,song2024microscopic,song2025electrically}. 

While the polariton landscape in monolayer TMDs is well-established \cite{chen2017valley,zhang2021van}, TMD bilayers exhibit a more complex material system. They host both intralayer (Fig.~\ref{fig:1}(b)) and spatially separated interlayer excitons (Fig.~\ref{fig:1}(c)) \cite{rivera2015observation,ovesen2019interlayer}. Due to the charge separation, interlayer excitons possess a permanent out-of-plane dipole moment, rendering them highly sensitive to external tuning via electric fields \cite{leisgang2020giant,hagel2022electrical,tagarelli2023electrical,erkensten2023electrically,lopriore2025enhancing}, and giving rise to strong repulsive dipole-dipole interactions (Fig.~\ref{fig:1}(c)) \cite{datta2022highly,louca2023interspecies,erkensten2023electrically,sun2024dipolar}. When these different exciton species are coupled to a microcavity mode, they give rise to dipolaritons: multi-component hybrid states that inherit both the large oscillator strength of intralayer excitons \cite{schneider2018two,fitzgerald2022twist} and the dipolar interactions of interlayer excitons \cite{datta2022highly,louca2023interspecies,konig2023interlayer}. 
In the monolayer limit, polariton nonlinearities are primarily driven by the Coulomb exchange interaction \cite{erkensten2021exciton}, whereas in bilayers, dipole-dipole interactions become the dominant mechanism \cite{erkensten2023electrically,tagarelli2023electrical}. In both systems, these effects are further supplemented by phase-space filling \cite{levinsen2019microscopic,zhang2021van,datta2022highly,choo2024polaronic}. A fully microscopic model capable of disentangling these different contributions to experimental observables, such as density-dependent energy shifts, is still lacking.

In this work, we develop a predictive many-particle theory of polariton-polariton interactions in MoS$_2$ monolayers and naturally 2H-stacked homobilayers by combining the density matrix formalism with the Hopfield method \cite{konig2023interlayer,fitzgerald2024circumventing}. This approach allows us to unravel the different contributions to the density-dependent energy renormalizations and explore experimentally accessible tuning knobs. Assuming densities far below the Mott transition, we find that phase-space filling effects are negligibly small compared to fermionic exchange interactions in monolayers and dipole-dipole interactions in bilayers. In monolayers, the nonlinear response is highly sensitive to temperature, specifically depending on whether the majority of the thermalized polaritons occupy states inside or outside the lightcone. In bilayers, the polaritonic dipole-dipole interaction strength is governed by the underlying interlayer character, which can be tuned by an electric field and can even lead to a collapse of the Rabi splitting. Overall, our results highlight the role of excitonic and photonic components in polariton-polariton interactions, revealing that polariton nonlinearities can be controlled by cavity detuning, temperature, and light-matter coupling strength. This is of key importance for isolating the many-body physics behind the nonlinear energy shifts observed in experiments.

\section*{Results}
\subsection*{Nonlinear interactions in TMD monolayers}
\label{sec:res1}
\begin{figure*}[!t]
    \centering
    \includegraphics[width=1\textwidth]{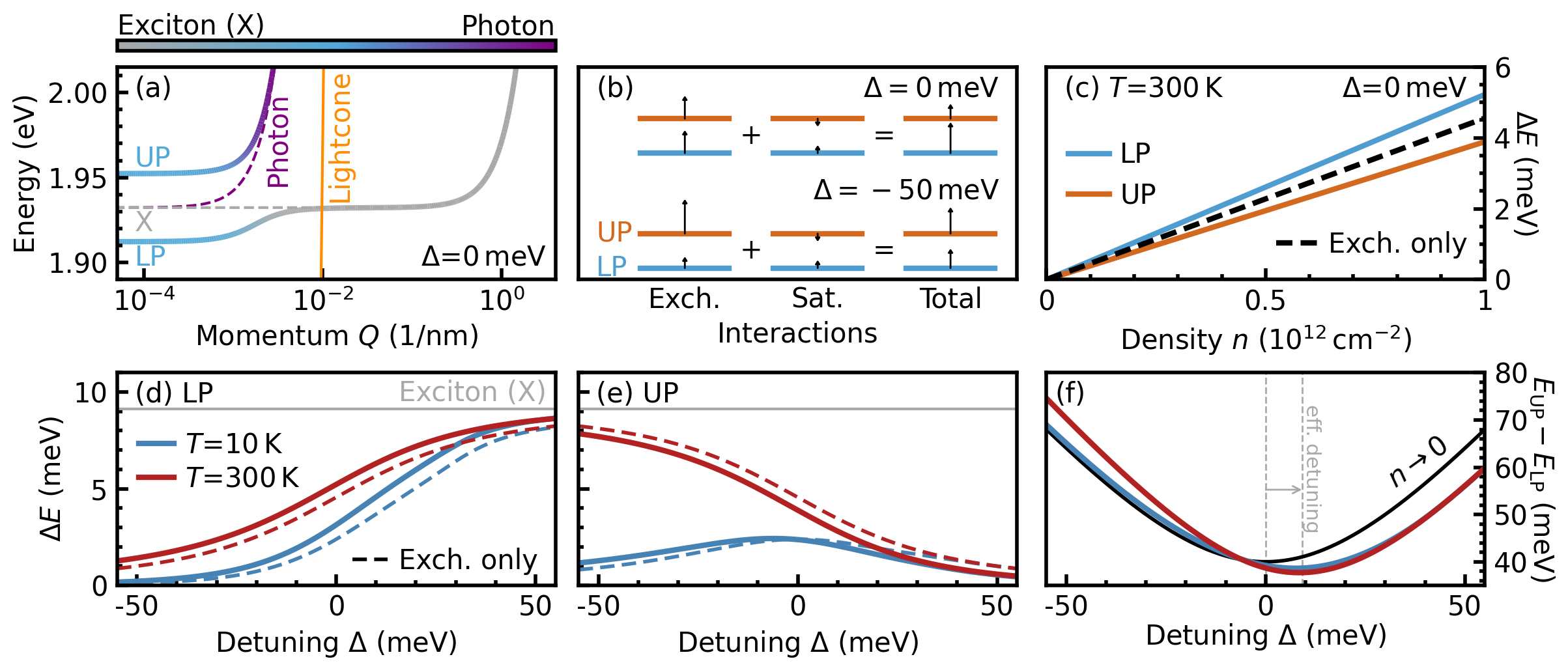}
    \caption{(a) Lower (LP) and upper (UP) polariton dispersions as a function of in-plane momentum $Q$ in the low-density limit at zero detuning ($\Delta=0$). The exciton/photon character is given by the Hopfield coefficients (color gradient). Gray and purple dashed lines show the bare exciton and cavity photon dispersions, respectively. The orange line denotes the lightcone edge. (b) Schematic of different LP and UP energy shift contributions due to nonlinear interactions for a zero- (top) and red- (bottom) detuned cavity. (c) The polariton energy shift $\Delta E$ is plotted against polariton density $n$ in monolayer MoS$_2$ at room temperature and zero detuning. The dashed line indicates the shift solely from exchange interaction, which is identical for the LP and UP at $\Delta=0$. (d) LP and (e) UP polariton energy shift versus detuning at a fixed polariton density of $n=10^{12}\,\mathrm{cm}^{-2}$ shown for cryogenic (blue) and room (red) temperatures. The horizontal gray line marks the temperature-independent shift in the bare exciton limit, which is purely due to the exchange interaction. (f) The resulting LP-UP energy splitting at the two considered temperatures. While the low-density limit (black line) exhibits a minimum at zero detuning, the inclusion of nonlinear interactions at an elevated density shifts the minimum to positive detunings, corresponding to an effective interaction-induced exciton detuning.
    }
    \label{fig:2}
\end{figure*}

At elevated excitation densities, the strong nonlinear interactions between polaritons cause a renormalization of the energy. For a TMD monolayer integrated within a Fabry-Pérot (FP) microcavity (Fig.~\ref{fig:1}(a)), the relevant mechanisms are the Coulomb exchange interaction and excitonic phase-space filling (saturation) both due to the fermionic substructure of the constituent excitons, as illustrated in Fig.~\ref{fig:1}(b). Microscopically, saturation arises from Pauli blocking of the fermionic substructure, which reduces the number of available electron–hole configurations and therefore the effective exciton oscillator strength. In recent experimental studies, it has been assumed for simplicity that the lower (LP) and the upper polariton (UP) branches exhibit density-induced energy shifts of the same magnitude \cite{zhang2021van,datta2022highly}. The exchange interaction typically causes a blue-shift of both branches, whereas the saturation leads to a reduction of the Rabi splitting, i.e., a blue-shift for the LP and a red-shift for UP. Our microscopic Wannier-Hopfield approach \cite{fitzgerald2022twist,konig2023interlayer} enables us to disentangle competing mechanisms for individual polariton branches and explore the impact of detuning and exciton-photon coupling strength. Using an equation-of-motion approach based on a many-body Hamiltonian including the Coulomb interaction and electron-photon interaction \cite{erkensten2023electrically,fitzgerald2022twist}, we obtain for the energy shift, $\Delta E_\nu$, of the polariton branch $\nu$ at normal incidence:
\begin{align}\label{eq:1_monolayer_shift}
\begin{split}
    \Delta E_{\nu}=&W |U_X^\nu(0)|^2\sum_{\nu',\bm{Q}}|U_X^{\nu'}(\bm{Q})|^2N_{\nu',\bm{Q}}\\
    &-g_\text{e}\sum_{\nu',\bm{Q}}\left(R_1^{\nu,\nu'}(\bm{Q})+R_2^{\nu,\nu'}(\bm{Q})\right)N_{\nu',\bm{Q}}\;\text{,}
    \end{split}
\end{align}
where $N_{\nu,\bm{Q}}$ is the polariton occupation with a 2D center-of-mass momentum $\bm{Q}$, and $U_{X(C)}^{\nu}(\bm{Q})$ are the respective excitonic (photonic) Hopfield coefficients that describe the hybrid character of the polariton. The first term in Eq.~(\ref{eq:1_monolayer_shift}) arises from the exchange interaction between the constituent electrons and holes within excitons. The second term stems from the saturation interaction associated with the fermionic substructure of the excitonic component of the polaritons, where $R_1^{\nu,\nu'}(\bm{Q})=R_{\bm{Q}} U_C^\nu(0)U_X^{*\nu}(0)|U_X^{\nu'}(\bm{Q})|^2$ and $R_2^{\nu,\nu'}(\bm{Q})=R_{\bm{Q}} U_C^{\nu'}(\bm{Q})U_X^{*\nu'}(\bm{Q})|U_X^\nu(0)|^2$. These describe the reciprocal saturation effects between polaritons inside the lightcone and the dark exciton reservoir. Here, $W$ and $R_{\bm{Q}}$ denote the exchange and saturation matrix element in the exciton basis respectively, while $g_\text{e}$ is the electron-photon coupling element. Note that the exchange-induced shift depends only on excitonic fractions, whereas the saturation-induced shift requires a certain mix of excitonic and photonic character. Further details on the derivation of Eq.~(\ref{eq:1_monolayer_shift}) can be found in section~S1.3 and S1.4 the supplementary information (SI).

In Fig.~\ref{fig:2}(a), we show the polariton energy landscape as a function of the in-plane momentum $Q$ at zero detuning ($\Delta = 0$) for a representative MoS$_2$ monolayer embedded in a FP cavity. Only the 1s exciton at the K-point in the Brillouin zone is considered. We chose a realistic exciton-photon coupling strength, yielding a Rabi splitting of 40\,meV \cite{dufferwiel2015exciton}. For this coupling strength, the occupation of the UP is negligible in thermal equilibrium at all temperatures. Consequently, only the LP branch ($\nu'=\mathrm{LP}$) contributes to the first term of Eq.~(\ref{eq:1_monolayer_shift}). As indicated by the color gradient in Fig.~\ref{fig:2}(a), at zero detuning and normal incidence ($Q = 0$), both polariton branches consist of 50\% excitonic and 50\% photonic character, i.e., $U_X^{\mathrm{LP}}(0)=U_X^{\mathrm{UP}}(0)$. Since both branches share the same excitonic character inside the lightcone, they experience an identical blueshift stemming from the exchange interaction. In contrast, the saturation induces a symmetric shift with respect to the exciton energy, i.e., blue-shifting the LP and red-shifting the UP. Overall, this leads to a slightly larger blueshift for the LP, as illustrated in the top schematic of Fig.~\ref{fig:2}(b). This is also visible in Fig.~\ref{fig:2}(c), where the exchange interaction produces identical energy shifts for the LP and UP (black dashed line), whereas the saturation enhances the LP blueshift (blue line) and suppresses the UP blueshift (orange line). It has been assumed in recent studies that the rigid exchange-induced shifts of the polariton branches extend to detuned cavities \cite{zhang2021van,datta2022highly} as well. However, these shifts are distinct for each branch, governed by their specific excitonic character. The polariton branch with the larger excitonic character experiences the greater energy shift: for a red-detuned cavity, the UP shifts more (lower schematic in Fig.~\ref{fig:2}(b)), whereas for a blue-detuned cavity, the larger shift occurs for the LP (see Fig.~S1 the SI for the Hopfield coefficients as a function of detuning).

The interaction-induced shift depends only on the LP occupation and the excitonic overlap of each branch with the LP. Fixing the polariton density at $10^{12}$\,cm$^{-2}$ and examining the energy shifts as a function of detuning (Figs.~\ref{fig:2}(d) and (e)), we find that for a red-detuned cavity ($\Delta<0$) the UP blueshifts more than the LP across all temperatures. This occurs because, at $Q=0$, the UP has a higher excitonic fraction, while the LP branch is predominantly photonic (see Fig.~S1 the SI), leading to a stronger exchange interaction for the UP branch. At high temperatures, the thermal energy exceeds the energy separation between the LP at $Q=0$ and the exciton energy, redistributing polariton occupation toward high-momentum reservoir states outside the lightcone, where the branch is predominantly excitonic. Similarly, the UP is mostly excitonic at $Q=0$ in a red-detuned cavity, therefore the energy shift approaches the excitonic limit. This behavior is illustrated in Fig.~\ref{fig:2}(e), where the red curve converges to the excitonic limit (horizontal grey line), as $\Delta$ becomes increasingly negative. Both the UP and the LP energy shifts are considerably reduced at low temperatures. Here, the LP occupation is concentrated within the lightcone due to the large energy separation between the LP and the bare exciton for a red-detuned cavity. The detuning-induced increase of the photonic character of the occupied LP states weakens the excitonic exchange interaction, thereby reducing the blueshifts for both branches, with the UP still shifting slightly more due to its larger excitonic character. Experimentally, larger energy shifts for the UP in the red-detuned case have been observed, while the overall shifts remain small at low temperatures, typically on the order of 1–2\,meV \cite{zhang2021van}.

 In the opposite case of a blue-detuned cavity ($\Delta>0$), the LP and UP are predominantly excitonic and photonic, respectively, across all momenta. Accordingly, the exchange shift for the LP approaches the excitonic limit (gray line in Fig.~\ref{fig:2}(d)), while for the UP it diminishes towards zero at all temperatures. Since the saturation interaction requires a hybrid excitonic and photonic character, this contribution to the energy shift becomes smaller for both large negative and positive detunings, where polaritons become either purely excitonic or photonic. The saturation effect can be identified by the difference between the solid and dashed lines in Figs.~\ref{fig:2}(d) and (e). For the LP, saturation always leads to an additional blueshift, with the dashed lines lying below the solid ones. For the UP, saturation generally results in a redshift, indicated by the dashed lines lying above the solid ones, except in the case of a strongly red-detuned cavity at low temperatures, where the UP instead exhibits a saturation-induced blueshift due to the relative phases of the relevant Hopfield coefficients, see section~S1.3 in the SI for details. Overall, despite the detuning- and temperature-dependent variations, the exchange interaction remains the dominant contribution to the energy shifts in TMD monolayers.

Next, we examine how these interaction-induced shifts modify the energy splitting between the LP and UP branches, shown in Fig.~\ref{fig:2}(f). The black curve represents the low-density limit ($n \rightarrow 0$), where the LP–UP separation exhibits the expected level-repulsion behavior: quadratic around zero detuning and linear at large detuning (see section S1.2 in the SI). The colored curves at finite density are obtained directly from the shifts shown in panels (d) and (e). Specifically, the modified splitting is determined by the bare separation plus the difference in the shift between the branches ($\Delta E_{\mathrm{UP}} - \Delta E_{\mathrm{LP}}$). Thus, deviations from the black curve reflect the imbalance between the interaction-induced shifts of the UP and LP. For negative detuning ($\Delta<0$), the UP blueshift shown in panel (e) is much larger than the corresponding LP blueshift in (d) for room temperature (red curves). Consequently, this leads to an enhanced LP–UP splitting, and the red curve in (f) lies above the black low-density reference. At cryogenic temperatures, where the UP and LP shifts are comparably small, the blue curve in (f) remains close to the black line. Around zero detuning, where the difference between the LP and UP shifts is minimal for both temperatures, the finite-density curves in (f) closely follow the black reference, aside from a small horizontal offset. In the exciton picture, this offset arises from the exchange-induced blueshift of the exciton resonance, which effectively detunes it from the cavity mode. In the polariton picture, this manifests as a horizontal translation of the LP and UP branches relative to the low-density case (dashed grey vertical lines in Fig.~\ref{fig:2}(f)). For positive detuning ($\Delta>0$), the behavior is reversed: the LP blueshift in panel (d) increases while the UP blueshift decreases in (e). Subtracting a larger LP shift while adding a smaller UP shift reduces the LP–UP splitting, causing both the red and blue curves in (f) to lie below the black reference. In this regime, the difference between temperatures is small, reflecting the similar behavior of the LP and UP shifts in panels (d) and (e).

\subsection*{Nonlinear interactions in TMD homobilayers subject to an electric field}
We next consider a homobilayer composed of naturally 2H-stacked MoS$_2$ \cite{liu2014evolution}, where electrons and holes possess an additional layer degree of freedom. We therefore introduce the compound index $L = (l_e, l_h, s)$, denoting electron layer, hole layer, and spin, respectively. This allows for intralayer excitons (X), in which the electron and hole reside in the same layer, as well as interlayer excitons (IX), where they occupy adjacent layers. The latter possess a negligible oscillator strength due to the spatial separation of the electron and hole \cite{ovesen2019interlayer,jiang2021interlayer}. X and IX states are coupled via hole tunneling between the layers, giving rise to hybrid excitons (hX) \cite{brem2020hybridized,hagel2022electrical}. These states acquire oscillator strength through their intralayer component and simultaneously exhibit a static out-of-plane electric dipole moment originating from their interlayer character \cite{konig2023interlayer}. This dipolar nature enables coupling to an out-of-plane electric field, leading to a tunable quantum-confined Stark shift \cite{leisgang2020giant,erkensten2023electrically,lopriore2025enhancing}. In the absence of an external electric field, the spin-up and spin-down configurations of each excitonic species are degenerate in homobilayers.
\begin{figure}[!t]
    \centering
    \includegraphics[width=.5\textwidth]{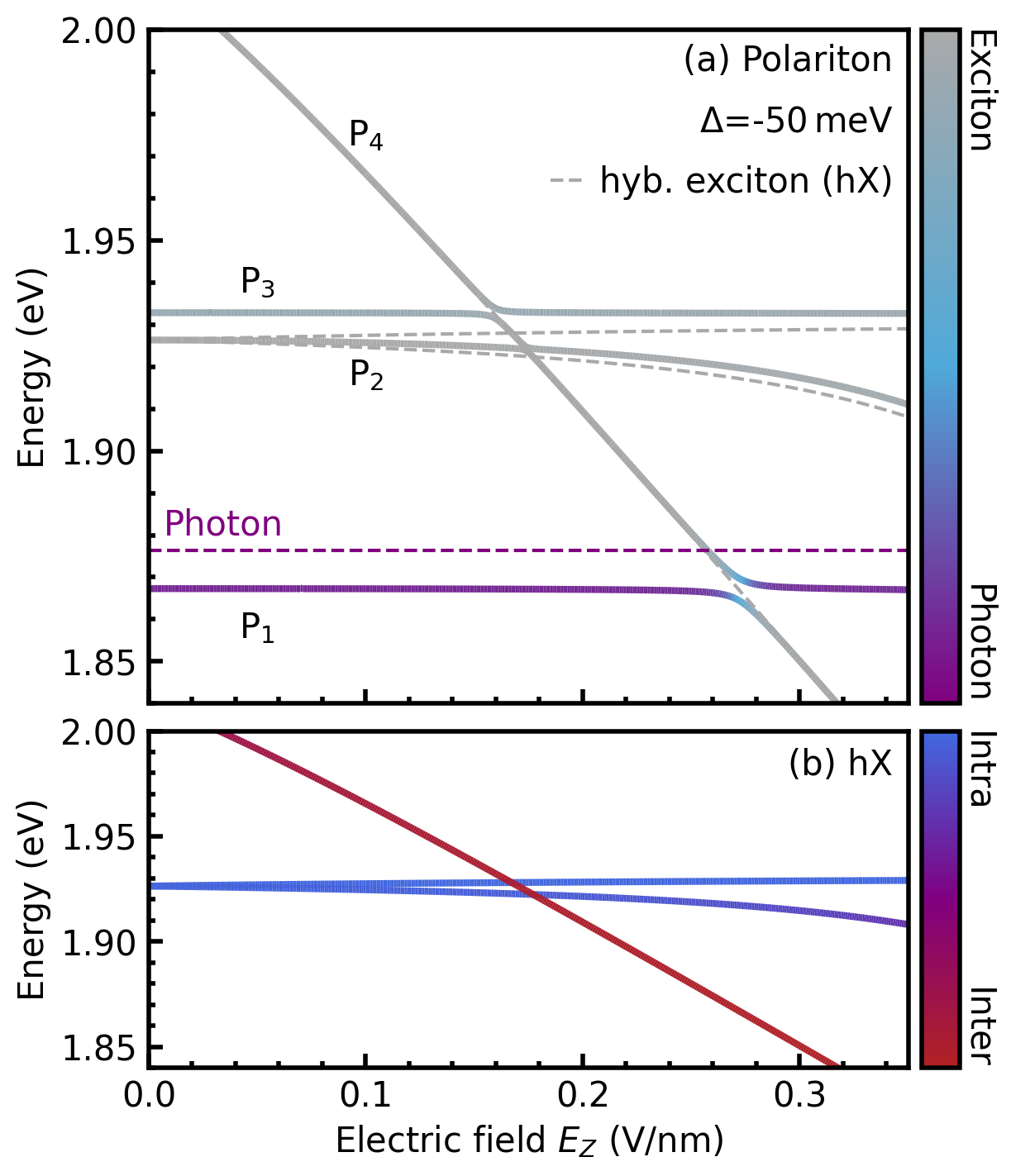}
    \caption{(a) Polariton energy landscape in a 2H-stacked MoS$_2$ homobilayer as a function of an applied out-of-plane electric field $E_\text{z}$ for a detuning of $\Delta=-50\,\mathrm{meV}$. The solid lines denote the four lowest-energy polariton branches, $P_1$ to $P_4$, with the color gradient showing the excitonic/photonic character. Gray and purple dashed lines indicate the hybrid exciton branches and the cavity photon energy, respectively. (b) Hybrid exciton energies from (a) with the color gradient showing their intra- vs. interlayer character. The energy shifts are driven by the interlayer component, which couples to the electric field due to the out-of-plane dipole moment.}
    \label{fig:3}
\end{figure}

Within the polariton picture, obtained by mixing hybrid excitons and the cavity photon mode through the Hopfield transformation \cite{konig2023interlayer}, we find that polariton energy shifts arising from both saturation and dipole–dipole interactions are negligible in the absence of an applied electric field. Saturation effects are even weaker than for the monolayer due to the reduced oscillator strength of IX states. In principle, repulsive dipole-dipole interactions give rise to large energy shifts \cite{lopriore2025electrically}, however, in a homobilayer, each interlayer exciton has a degenerate counterpart with the opposite layer configuration and thus an opposite dipole moment. As a result, the net interaction is canceled out \cite{erkensten2023electrically}. This degeneracy can be lifted by an external electric field via the Stark effect, allowing net dipolar interactions to emerge. The electric field dependence of the polariton dispersions is directly inherited from the constituent hybrid excitons \cite{erkensten2023electrically,lopriore2025electrically,song2024microscopic}. To ensure a realistic light–matter coupling, we set the oscillator strength of the lowest lying, intralayer-like hybrid exciton to give a zero-field Rabi splitting of 40\,meV \cite{louca2023interspecies}, accounting for the intrinsic two-fold degeneracy of the excitonic state. We define zero detuning with respect to this lowest-lying hybrid exciton at zero field.

We first focus on the four lowest-energy polariton branches P$_1$-P$_4$ for a red-detuned cavity ($\Delta=-50\,\mathrm{meV}$) at normal incidence in the low-density regime without any nonlinear interaction present. We find that the Rabi splittings for field strengths less than 0.2\,V/nm are very small, about 2.5\,meV at 0.16\,V/nm and 1\,meV at 0.175\,V/nm (Fig.~\ref{fig:3}(a)), due to the cavity photon being energetically much lower. The resulting polariton band structure is dictated by the underlying hybrid-exciton landscape \cite{sponfeldner2022capacitively,lopriore2025electrically}, shown as grey dashed lines in Fig.~\ref{fig:3}(a). Owing to its static out-of-plane dipole moment, the IX constituent experiences a linear shift under an out-of-plane electric field, while the X constituent remains principally unaffected \cite{leisgang2020giant,hagel2022electrical}. This is directly reflected in the hybrid exciton landscape (Fig.~\ref{fig:3}(b)). While the strongly intralayer-like excitons (blue gradient) remain relatively flat, the interlayer-like hybrid excitons (red gradient) exhibit an almost linear shift. At a field strength larger than $0.17\,\mathrm{V/nm}$, a predominately interlayer-like hybrid exciton becomes the lowest lying state. The hybridization of this state in the range of fields shown in Fig.~\ref{fig:2} is low, because the underlying interlayer exciton IX$_{(l_e=1,l_h=0,s=0)}$ is coupling to the intralayer exciton X$_{l_e=1,l_h=1,s=0}$ (B-exciton) via hole tunneling, but the large energy separation between the two states, arising from the strong spin-orbit splitting in the valence band \cite{kormanyos2015k}, suppresses the hybridization \cite{konig2023interlayer}. Furthermore, the IX$_{(l_e=0,l_h=1,s=0)}$ approaches the intralayer excitons X$_{l_e=1,l_h=1,s=0}$ and X$_{l_e=1,l_h=1,s=1}$ (A and A' exciton) near 1.93\,eV. The interlayer exciton could, in principle, couple to the former via electron tunneling, however, this process is symmetry-forbidden in a 2H-stacked homobilayer \cite{ruiz2019interlayer,hagel2021exciton}. Coupling to the A' exciton is also prohibited due to their opposite spin configurations. As a result, the hybridization remains weak throughout the electric-field and energy ranges shown in Fig.~\ref{fig:3}(b). The oscillator strength is therefore small and the Rabi splitting found at about 0.27\,V/nm in Fig.~\ref{fig:3}(a) is only 5\,meV. In summary, for a strongly red-detuned cavity at zero-field, the intralayer-like excitons possess a large oscillator strength, but they are energetically far from the photon mode and exhibit a negligible tunability under an electric field. Conversely, the much more tunable interlayer-like excitons have a much weaker oscillator strength, resulting in a small Rabi splitting. Results for zero- and blue-detuned regimes are presented in Fig.~S3 in the SI.

We now turn to the nonlinear regime in the TMD homobilayer and include the dipole-dipole repulsion as well as the saturation interaction, as illustrated in Fig.~\ref{fig:1}(c). Intralayer exchange interactions are known to provide only minor corrections to the direct dipole–dipole repulsion \cite{erkensten2023electrically,steinhoff2024exciton} and have a negligible impact on experimentally accessible density-dependent energy renormalizations \cite{lopriore2025electrically,federolf2025tuning}. This is because their contributions are largely canceled by higher-order correlation effects \cite{katsch2019theory,trovatello2022disentangling}. We again employ the equation-of-motion approach to obtain the hybrid exciton polariton energy shifts at elevated densities:
\begin{align}
    \begin{split}
    \Delta E_\nu=&\sum_{\bm{Q},\nu'}\left(\mathcal{D}^{\nu\nu'}_1(\bm{Q})+\mathcal{D}^{\nu\nu'}_2(\bm{Q})\right) N_{\nu',\bm{Q}}\\
    &-\sum_{\bm{Q},\nu'}g_\text{e}\left(\mathcal{R}^{\nu\nu'}_1(\bm{Q})+\mathcal{R}^{\nu\nu'}_2(\bm{Q})\right) N_{\nu',\bm{Q}}
    \end{split}\;\text{,}\label{eq:1_bilayer_shift}
\end{align}
with the dipole-dipole (first line) and saturation (second line) interactions determined by the momentum-dependent matrix elements 
$\mathcal{D}^{\nu\nu'}_1(\bm{Q})$, $\mathcal{D}^{\nu\nu'}_2(\bm{Q})$ and $\mathcal{R}^{\nu\nu'}_1(\bm{Q})$, $\mathcal{R}^{\nu\nu'}_2(\bm{Q})$, respectively. 
More details are given in section~S1.3 and S1.5 in the SI.

\begin{figure}[!t]
    \centering
    \includegraphics[width=0.5\textwidth]{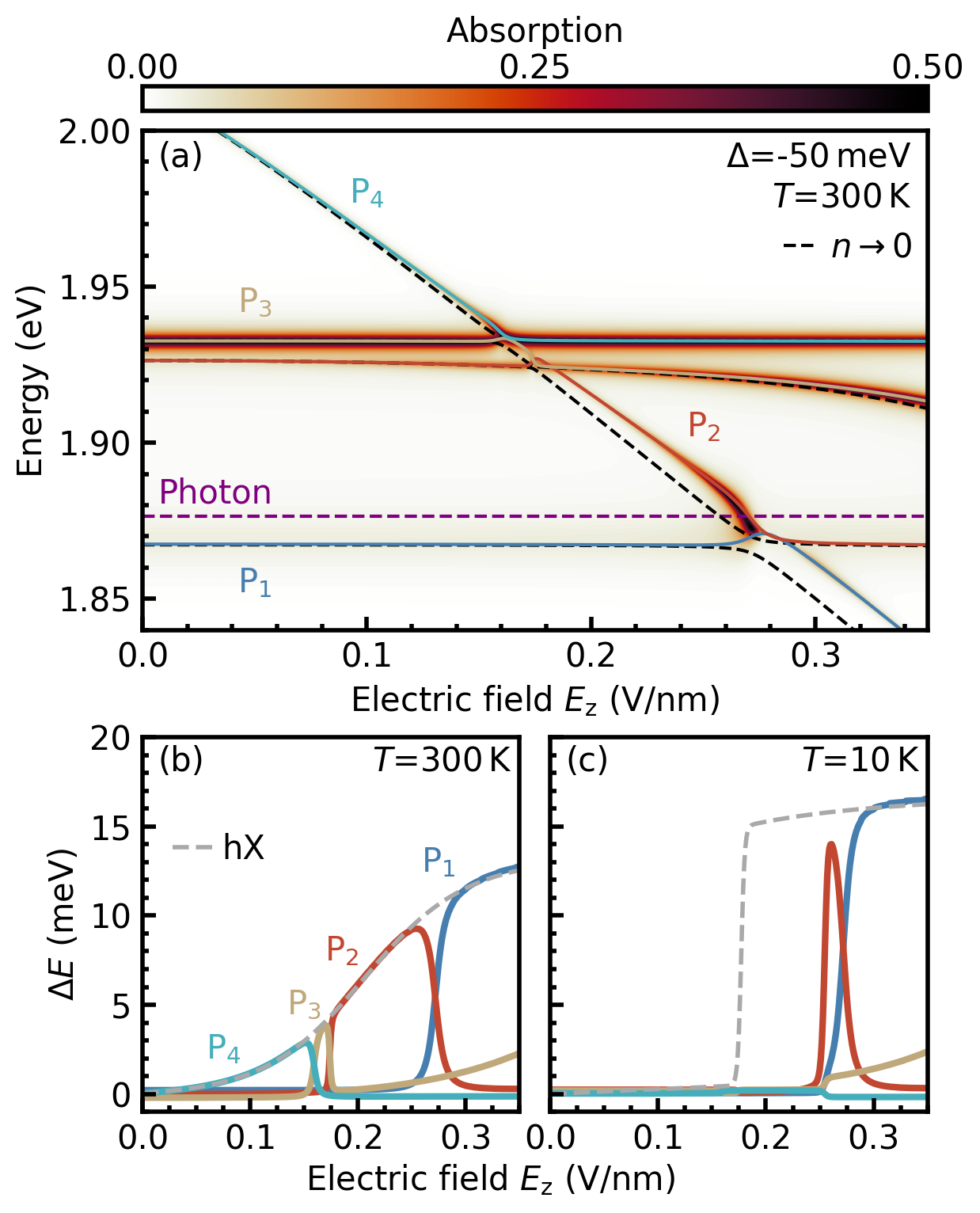}
    \caption{(a) Impact of interaction-induced shifts on the polariton energy landscape in a 2H-stacked MoS$_2$ homobilayer under an applied out-of-plane electric field $E_\text{z}$. Results are shown for the four lowest polariton branches at a detuning of $\Delta=-50\,\mathrm{meV}$, a temperature of $T=300\,\mathrm{K}$, and a polariton density of $n=10^{12}\,\mathrm{cm^{-2}}$. The colormap shows the polariton absorption. The low-density limit, where the interaction-induced shifts are zero, is denoted by the black dashed lines. (b)-(c) Field-dependent nonlinear energy shifts resolved for different polariton branches at (b) 300\,K and (c) 10\,K. The gray dashed line illustrates the shift of the predominantly interlayer-like hybrid exciton.}
    \label{fig:4}
\end{figure}

To assess the nonlinear energy shifts, we fix the polariton density to 10$^{12}$\,cm$^{-2}$, where the resulting interaction-induced shifts are small compared to the quantum-confined Stark shift induced by the electric field, as seen from the modest separation between the solid lines and the low-density regime (denoted by the dashed lines) in Fig.~\ref{fig:4}(a). Saturation effects are even less significant than in the monolayer case. This is because, in the polariton picture, a reduced oscillator strength translates into smaller Rabi splittings and consequently weaker nonlinear energy shifts. As expected from dipole–dipole interactions, sizable interaction-induced shifts occur only when the given polariton branch acquires a substantial interlayer-like character. These energetic shifts are still sufficient to close the Rabi-splitting under appropriate electrical tuning, as seen between the branches P$_1$ and P$_2$ at 0.27\,V/nm in Fig.~\ref{fig:4}(a). As the bare exciton energy crosses the fixed photon mode, the polariton branches exchange their excitonic character: the lower branch is more photonic before the avoided crossing and becomes more excitonic after, while the upper branch does the opposite. Since the nonlinear shift mainly stems from the interlayer-like excitonic component and scales with the corresponding Hopfield coefficient (see Eq.~(\ref{eq:1_bilayer_shift}) and Fig.~S4 in the SI), it effectively follows the excitonic fraction of each branch. This reduces the effective level repulsion when the interlayer-like hybrid exciton is brought into resonance with the photon and can partially or even completely close the avoided crossing. Consequently, the absorption, shown by the background color map, follows these shifted polariton branches, making them readily observable experimentally. Notably, the absorption feature at the avoided crossing exhibits a distinct profile: the absorption along the polariton branch stops abruptly at about 0.27\,V/nm, contrasting with typical low-density spectra (see Figure S6 in the SI).

Examining the isolated contributions to the interaction-induced polariton energy shift in Fig.~\ref{fig:4}(b), we find that at any given electric field, the underlying interlayer-like hybrid exciton (dashed grey line) predominantly contributes to the shift of one polariton branch, with the dominant branch depending on the field strength. Specifically, the nonlinear shift follows the excitonic character, so the polariton branch closest to the hybrid exciton acquires a shift equal to that of the exciton at the same density, and this shift is partitioned between branches as their excitonic content is exchanged. A similar trend appears in Fig.~\ref{fig:4}(c), however, at cryogenic temperatures the occupation is concentrated in the lowest-lying polariton branch P$_1$. Due to the red-detuning, P$_1$ is even lower than the lowest exciton state. Therefore, a larger electric field is required for the interlayer-like polariton branch to reach this occupied energy region, compared to higher temperatures where mostly states in the excitonic reservoir are occupied. As a consequence, the entire nonlinear response is shifted toward higher electric fields compared to the hybrid-exciton case (dashed grey line). Electrically tunable dipolar polaritons were recently realized experimentally \cite{xiang2026electrically}, where the observed polariton energy shifts are comparable to those predicted in our study. In particular, at zero field the shifts are much smaller, in agreement with our findings. For other detuunings regimes see Fig.~S5 in the SI.

\subsection*{Tuning the electron-photon coupling strength}
\begin{figure}[!t]
    \centering
    \includegraphics[width=0.5\textwidth]{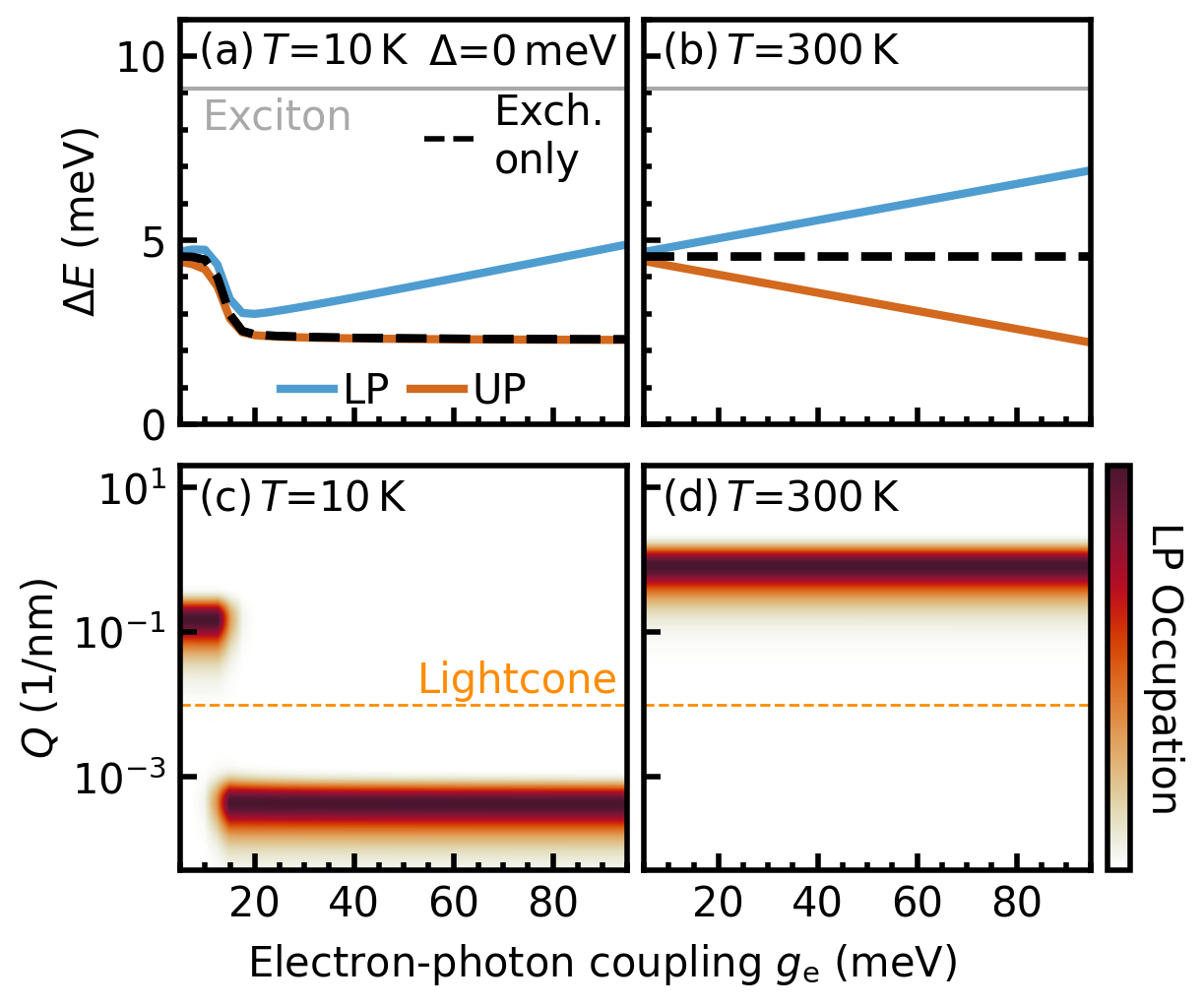}
    \caption{Nonlinear polariton energy shifts in monolayer MoS$_2$ embedded within a FP cavity at (a) $T=10\,K$ and (b) $T=300\,K$ for zero detuning as a function of the electron-photon coupling strength $g_\text{e}$. The dashed line denotes the shift due to just exchange interaction, which is symmetric for the LP and UP branches. The bare exciton energy shift is marked by the horizontal solid gray line. LP occupation vs. $g_\text{e}$ and momentum $Q$ at (c) $T=10\,K$ and (d) $T=300\,K$. The horizontal line indicates the edge of the lightcone. At low temperatures and coupling strengths of above 15\,meV most of the polaritons are concentrated within the lightcone, leading to the sharp reduction in the exchange interaction shift observed in (a).}
    \label{fig:5}
\end{figure}

We now explore how tuning the electron–photon coupling, $g_\text{e}$, modifies the relative contributions to the polariton energy shifts at elevated densities. Exchange- and dipole-driven interactions are controlled solely by the excitonic character of the polariton, entering through the Hopfield coefficients and the resulting state occupations. In contrast, saturation-induced shifts show an additional linear dependence on the coupling strength itself, shown in Eqs.~(\ref{eq:1_monolayer_shift}) and (\ref{eq:1_bilayer_shift}), allowing the two mechanisms to be selectively tuned relative to each other. We restrict our analysis to monolayer MoS$_2$ at zero detuning (see Figs.~S7 and S8 in the SI for the corresponding homobilayer results). We first focus on the exchange contribution to the energy shifts, as shown by the dashed line in Fig.~\ref{fig:5}(a) and (b). As already shown in Fig.~\ref{fig:2}(c), the LP and UP branches exhibit identical exchange-induced shifts at zero detuning. The dependence of the shift on the coupling strength is strongly influenced by temperature: while for room temperature the shift is independent of $g_\text{e}$, at cryogenic temperatures we find a step-like decrease at approximately $g_\text{e}=15\,\mathrm{meV}$. This behavior can be traced back to Eq.~(\ref{eq:1_monolayer_shift}), where the exchange shift explicitly depends on the occupation of polariton states. Owing to the large splitting between the LP and UP branches, only the LP states are significantly populated across the considered temperature and coupling strength ranges. For weak coupling strengths, most particles reside outside the lightcone even at low temperatures, see Fig.~\ref{fig:5}(c). Increasing the coupling strength above approximately 15\,meV lowers the LP energy sufficiently to concentrate the population within the lightcone. At room temperature, however, the majority of polaritons lie outside of the lightcone for all considered $g_\text{e}$ due to the low density of states within the lightcone. This dependence of the occupation on detuning and temperature directly affects the exchange shift. When the fully excitonic LP outside the lightcone interacts with a 50\% excitonic LP or UP inside the lightcone, the exchange shift is halved relative to the fully excitonic limit. This explains the halved exchange shift at low $g_\text{e}$ compared to the excitonic case (grey solid line) in Fig.~\ref{fig:5}(a). When 50\% excitonic LP polaritons inside the lightcone interact with 50\% excitonic LP or UP polaritons, the shift is further reduced to one quarter of the fully excitonic value, as observed in Fig.~\ref{fig:5}(a) for coupling strengths above approximately 15\,meV. At room temperature, the exciton reservoir is highly populated (Fig.~\ref{fig:5}(d)), regardless of the coupling strength. As a result, the exchange shift is constant as a function of coupling strength and about half of the excitonic limit, see Fig.~\ref{fig:5}(b).

For the saturation contribution, the underlying microscopic mechanisms are more complex: it depends not only on the excitonic Hopfield coefficients, but also on the overlap between excitonic and photonic components. The resulting saturation shift scales overall linearly with $g_\text{e}$. Its magnitude, however, is determined by the momentum distribution of the LP population. In practice, the occupation resides either in the excitonic reservoir at large momenta or at small but finite momenta within the light cone (since the density of states vanishes at $Q=0$). The shift is thus controlled by the Hopfield coefficients at precisely these momenta, which at zero detuning are essentially independent of the coupling strength (see Fig.~S2 in the SI). Owing to the overall linear scaling with $g_\text{e}$, the saturation contribution remains negligible for $g_\text{e}\leq 15,\mathrm{meV}$ compared to the exchange-induced shift. At low temperatures, both terms of the saturation interaction in Eq.~(\ref{eq:1_monolayer_shift}) must be considered. In the first term, $R_1^{\nu,\mathrm{LP}}$, the excitonic part of the LP, $|U_X^\mathrm{LP}(\bm{Q})|^2$, is multiplied by the overlap of excitonic and photonic Hopfield coefficients of the branch $\nu$, i.e., $U_C^\nu(0)U_X^{*\nu}(0)$. In the second term, $R_2^{\nu,\mathrm{LP}}$, the overlap of excitonic and photonic components of the LP, $U_C^\mathrm{LP}(\bm{Q})U_X^{*\mathrm{LP}}(\bm{Q})$, is multiplied by the excitonic fraction of the branch $\nu$, $|U_X^\nu(0)|^2$. At cryogenic temperatures, most particles occupy states within the lightcone, where both branches consist of approximately 50\% excitons and 50\% photons. This means that all magnitude factors in these terms are close to 0.5. The crucial difference lies in the \emph{relative phase} of the Hopfield coefficients: for the LP, the excitonic and photonic components have the same sign, while for the UP they have opposite signs (see section S1.2 in the SI for details). As a result, the two terms for LP add constructively, $R_1^\mathrm{LP,LP} = R_2^\mathrm{LP,LP}$
producing a net positive linear shift with the coupling strength, as seen by the increasing difference between the black-dashed and blue line in Fig.~\ref{fig:5}(a). For the UP, the two terms cancel, $-R_1^\mathrm{UP,LP} = R_2^\mathrm{UP,LP}$, resulting in a vanishing shift (black-dashed and orange line in Fig.~\ref{fig:5}(a)).

At high temperatures, the situation changes: the LP is fully excitonic and mostly states outside the lightcone are occupied. Therefore, the overlap in the second term vanishes, $R_2^{\nu,\mathrm{LP}} = 0$, leaving only the first term. Due to the opposite phases of the excitonic–photonic overlap for the LP and UP, the first term now blueshifts the LP and redshifts the UP, shown by the linearly increasing difference between the black-dashed line, and the blue and orange lines, respectively, in Fig.~\ref{fig:5}(b). Taken together, examining the interaction-induced shifts as a function of the exciton–photon coupling strength shows that it can provide additional control over the degree of light–matter hybridization and which interaction channels dominate the nonlinear response.

\section*{Discussion}
Our comprehensive microscopic investigation of nonlinear energy shifts of exciton-polaritons in TMD monolayers and homobilayers offers critical insights that challenge current phenomenological models and advance the microscopic understanding of strong light-matter coupling. We have shown that interaction-induced shifts are asymmetric between the lower and upper polariton branches, as the exchange-induced shift is proportional to the excitonic fraction of the respective polariton state. Furthermore, temperature and exciton–photon coupling strength emerge as key quantities to control the nonlinear response. At low temperatures, the polariton occupation depends sharply on the coupling strength, exhibiting a transition from an excitonic reservoir outside the lightcone to states inside the lightcone. This transition accounts for the abrupt change in the exchange-driven shift observed at cryogenic temperatures and provides a tunable way to probe thermalization and condensation dynamics, since the nonlinear response directly reflects whether the population is equilibrated inside or outside the lightcone. Our study on TMD homobilayers exploits the effect of an out-of-plane electric field and the associated dipole-dipole interactions. We established that the required interaction-induced shifts for device functionality are critically dependent on the polariton state acquiring a significant interlayer-like character, allowing the permanent dipole moment to couple effectively to the electric field and give rise to efficient repulsive dipole-dipole forces. Crucially, the resulting shifts are sufficient to reduce the Rabi splittings to zero, highlighting a powerful route toward the all-electrical control of the polariton energy landscape and directional flow \cite{gu2021enhanced,genco2025femtosecond}. Overall, our work provides a microscopic framework for understanding and engineering nonlinear polaritons in TMD heterostructures. These findings are crucial for guiding future experimental investigations and represent a significant advance toward utilizing the unique nonlinear properties of 2D semiconductor polaritons for the development of dynamically controlled integrated photonic circuits.

\section*{Methods}
We model excitonic states in mono- and bilayer TMD systems using the Wannier equation within an effective-mass approximation \cite{kormanyos2015k}, incorporating a screened Coulomb interaction \cite{rytova1967screened,keldysh1979coulomb,ovesen2019interlayer,brem2020phonon}. Only electron–hole pairs at the K-points in the Brillouin zone can reside in the same valley and couple strongly to light \cite{brem2020phonon}, therefore we restrict the analysis to these valleys only. For the bilayer case, there are two possible layer indices and spin states for both electrons and holes. This gives a total of $2^4 = 16$ excitonic configurations. However, when focusing on optically bright excitons, the electron and hole must have the same spin \cite{feierabend2020brightening}, reducing the number of distinct electron–hole configurations to eight, giving four intra- and four interlayer excitons. Furthermore, interlayer hole tunneling between layers is included \cite{brem2020hybridized}. This approach provides full access to the excitonic energy landscape and corresponding wavefunctions. To describe the emergence of polaritonic states, we integrate the 2D semiconductor within a FP cavity and apply the Hopfield formalism to diagonalize the coupled exciton–photon Hamiltonian \cite{hopfield1958theory,fitzgerald2022twist}. The resulting Hopfield coefficients quantify the excitonic and photonic contributions to the polariton modes. In the monolayer case, these coefficients directly describe the admixture of the bare exciton and cavity photon mode. In contrast, for bilayer systems, hole tunneling leads to the formation of hybridized excitonic states, such that the polaritons inherit their matter character from these hybrid excitons. To resolve the composition of the polaritons in terms of the underlying bare excitons, we introduce generalized Hopfield coefficients that describe the photonic and bare intra-/interlayer exciton character \cite{konig2023interlayer}. To address nonlinear saturation effects, we consider the underlying electron–photon Hamiltonian \cite{fitzgerald2022twist}. From the resulting equation of motion for the microscopic polarization, we obtain the energy shifts associated with the saturation terms in Eqs.~(\ref{eq:1_monolayer_shift}) and (\ref{eq:1_bilayer_shift}), respectively.

Furthermore, we also include exciton-exciton interactions in the form of fermionic exchange interactions for TMD monolayers and dipole-dipole interactions for TMD bilayers, derived from a purely electron-hole Coulomb Hamiltonian \cite{erkensten2023electrically}. The resulting equations of motion are then transformed into the corresponding polariton basis, giving rise to the first term in Eqs.~(\ref{eq:1_monolayer_shift}) and (\ref{eq:1_bilayer_shift}), respectively. Lastly, the energy shifts due to these polariton-polariton interactions depend on the polariton occupation. In the thermalized limit, we describe the polariton populations by a Boltzmann distribution, using the low-density polariton dispersion and neglecting self-consistent feedback of the density on the spectrum or bottleneck effects \cite{fitzgerald2024circumventing}. We further assume fixed excitonic wavefunctions and restrict to density regimes where interaction-induced modifications of exciton wavefunctions and screening effects are expected to be small \cite{erkensten2023electrically,steinhoff2024exciton}. Absorption spectra are calculated using the polaritonic Elliott formula \cite{fitzgerald2022twist}. Further details on the theoretical approach can be found in section S1 the SI.

\bibliography{ref}
\section*{Acknowledgments}
    We thank Giuseppe Meneghini (Philipps-Universität Marburg) for his helpful contributions to the design and preparation of Figure 1. We acknowledge funding from the Deutsche Forschungsgemeinschaft (DFG) via the regular project 524612380.
\section*{Author contributions}
J.K.K. performed the calculations and simulations. J.K.K., J.M.F., D.E. and E.M. analyzed the obtained results. J.K.K. drafted the manuscript with all the authors contributing to the discussion and preparation of the paper. E.M. supervised the project.
\section*{Competing interests}
The authors declare no competing interests.
\section*{Data availability}
Data underlying the results presented in this paper are not publicly available at this time but may be obtained from the authors upon reasonable request.
\end{document}


\makeatletter
\renewcommand{\p@subsection}{}
\makeatother

\title{Supplementary Information\\Exciton Polariton–Polariton Interactions in Transition-Metal Dichalcogenides}
\author{Jonas K. K\"onig}
\affiliation{Department of Physics, Philipps-Universit{\"a}t Marburg, 
35037 Marburg, Germany}
\affiliation{mar.quest|Marburg
Center for Quantum Materials and Sustainable Technologies,
35032 Marburg, Germany}
\email{jonas.koenig@physik.uni-marburg.de}
\author{Jamie M. Fitzgerald}
\affiliation{Department of Physics, Philipps-Universit{\"a}t Marburg, 
35037 Marburg, Germany}
\affiliation{mar.quest|Marburg
Center for Quantum Materials and Sustainable Technologies,
35032 Marburg, Germany}
\author{Daniel Erkensten}
\affiliation{Department of Physics, Philipps-Universit{\"a}t Marburg, 
35037 Marburg, Germany}
\affiliation{mar.quest|Marburg
Center for Quantum Materials and Sustainable Technologies,
35032 Marburg, Germany}
\author{Ermin Malic}
\affiliation{Department of Physics, Philipps-Universit{\"a}t Marburg, 
35037 Marburg, Germany}
\affiliation{mar.quest|Marburg
Center for Quantum Materials and Sustainable Technologies,
35032 Marburg, Germany}
\maketitle
\section{Theoretical model}
\label{sec:theory}

\subsection{Wannier Equation and layer hybridization}
\label{subsec:Wannier_hyb}
To describe the excitonic optical response of mono- and homobilayer MoS$_2$, we employ the effective mass approximation and focus on the K points within the Brillouin zone. The exciton binding energies, $E_{b,n}^{L}$, and momentum-dependent wave functions, $\Psi_n^{L}(\bm{k})$, are obtained by solving the Wannier equation \cite{kira2006many} 
\begin{align}
   \sum_{\bm{k}'}\left(\frac{\hbar^2\bm{k}'^2}{2 m_r^{L}}\delta_{\bm{k}\bm{k}'}+V_{|\bm{k}-\bm{k}'|}^{L}\right)\Psi_n^{L}(\bm{k}')&=E_{b,n}^{L}(\bm{k})\Psi_n^{L}(\bm{k})\;\text{,}
   \label{eq:wannier_rel}
\end{align}
where the compound index $L=(l_e,l_h,s)$ denotes the exciton's electron/hole layer indices and the spin configuration \cite{ovesen2019interlayer}. For a given $L$, the solutions to the Wannier equation yield a hydrogen-like series of eigenstates characterized by the principal quantum number $n$, however, in this work we focus only on 1s-like excitons. In the bilayer case, we consider all possible layer and bright-spin configurations, whereas for the monolayer we include only the lowest-energy spin configuration. In Eq.~\ref{eq:wannier_rel}, $V_{|\bm{k}-\bm{k}'|}^{L}$ represents the screened Coulomb potential for the respective mono- or bilayer geometry \cite{ovesen2019interlayer,brem2020phonon}, while $m_r^{L}=\frac{m_e^{l_e,s}m_h^{l_h,s}}{m_e^{l_e,s}+m_h^{l_h,s}}$ is the reduced mass. The parameters for the effective electron (hole) mass $m^{l_{e(h)},s}_{e(h)}$ are taken from Ref.~\cite{kormanyos2015k}, while the dielectric constants in the Coulomb potential are from Ref.~\cite{laturia2018dielectric}. The total energy of excitons is then given by $E^{(X)}_{L,\bm{Q}}=\Delta +E_{b,n}^{L}+\delta E^L+\frac{\hbar^2 \bm{Q}^2}{2 M_L}$, where $\Delta$ denotes the bandgap and $\delta E^L$ the respective conduction- and valence band offsets due to the spin-orbit coupling. Here, $\bm{Q}$ denotes the center-of-mass momentum of the exciton, while $M^L=m_e^{l_e,s}+m_h^{l_h,s}$ is the total exciton mass. Here we set the band gap in such a way that the lowest lying exciton state has the same spectral position as observed in experiments \cite{louca2023interspecies}.

The excitonic eigenstates of the bilayer are characterized by a hybridization of layer-localized configurations driven by the spin-conserving tunneling of electrons and holes between the constituent layers. This coupling leads to the formation of hybrid excitons, which exist as a superposition of intra- and interlayer states. In the naturally occurring 2H stacking considered here, symmetry constraints dictate that electron tunneling is forbidden, leaving hole tunneling as the sole mechanism for hybridization \cite{ruiz2019interlayer,hagel2021exciton}. To include these processes into our model, we define an excitonic Hamiltonian where the hole tunneling acts as a coupling term between different layer-localized configurations. The hybrid exciton states are then determined by diagonalizing the following bosonic Hamiltonian:
\begin{align}
    \hat{H}_0=\sum_{\bm{Q},L} E_{L,\bm{Q}}^{(X)} \hat{X}_{L,\bm{Q}}^\dagger \hat{X}_{L,\bm{Q}}+\sum_{L\neq L',\bm{Q}}T_{LL'}\hat{X}_{L,\bm{Q}}^\dagger \hat{X}_{L',\bm{Q}}\;\text{,}
\end{align}
where $\hat{X}_{L,\bm{Q}}^{(\dagger)}$ is the excitonic annihilation (creation) operator. Here, $T_{LL'}$ describes the coupling between excitons with index $L$ and $L'$, and its form can be found in Ref.~\cite{hagel2022electrical}. Importantly, it describes the situation where two excitons share either the electron or the hole. Diagonalizing leads to the hybrid exciton Hamiltonian:
\begin{align}\hat{H}_0=\sum_{\eta,\bm{Q}}E_{\eta,\bm{Q}}^{(Y)}\hat{Y}_{\eta,\bm{Q}}^\dagger \hat{Y}_{\eta,\bm{Q}}\;\text{,}
\end{align}
where $E_{\eta,\bm{Q}}^{(Y)}$ is the energy of hybrid exciton $\eta$ and $\hat{Y}_{\eta,\bm{Q}}^{(\dagger)}$ is the respective annihilation (creation) operator. The mixing coefficients $C_L^\eta(\bm{Q})$ describing this basis transformation allow us to also express the coupling of hybrid excitons to photons as $g^{(Y)}_\eta(\bm{Q})=\sum_Lg^{(X)}_L C_L^\eta(\bm{Q})$, where $g^{(X)}_L=\sum_{\bm{k}}\Psi_\mathrm{1s}^L(\bm{k}) g_\text{e}$ and $g_\text{e}$ is the electron-photon coupling. The electric field effects are then included via a linear quantum-confined Stark-shift \cite{hagel2022electrical} for interlayer excitons $E_{L,\bm{Q}}^{(X)}\big|_{\text{Stark}}=-d_{(l,1-l,s)} E_\text{z}$, where $d_{L}$ is the out-of-plane dipole moment and $E_z$ is the electric field applied perpendicular to the material plane.

\subsection{Hopfield approach}
\label{subsec:Hopfield}
\begin{figure}[!t]
     \centering
     \includegraphics[width=.75\textwidth]{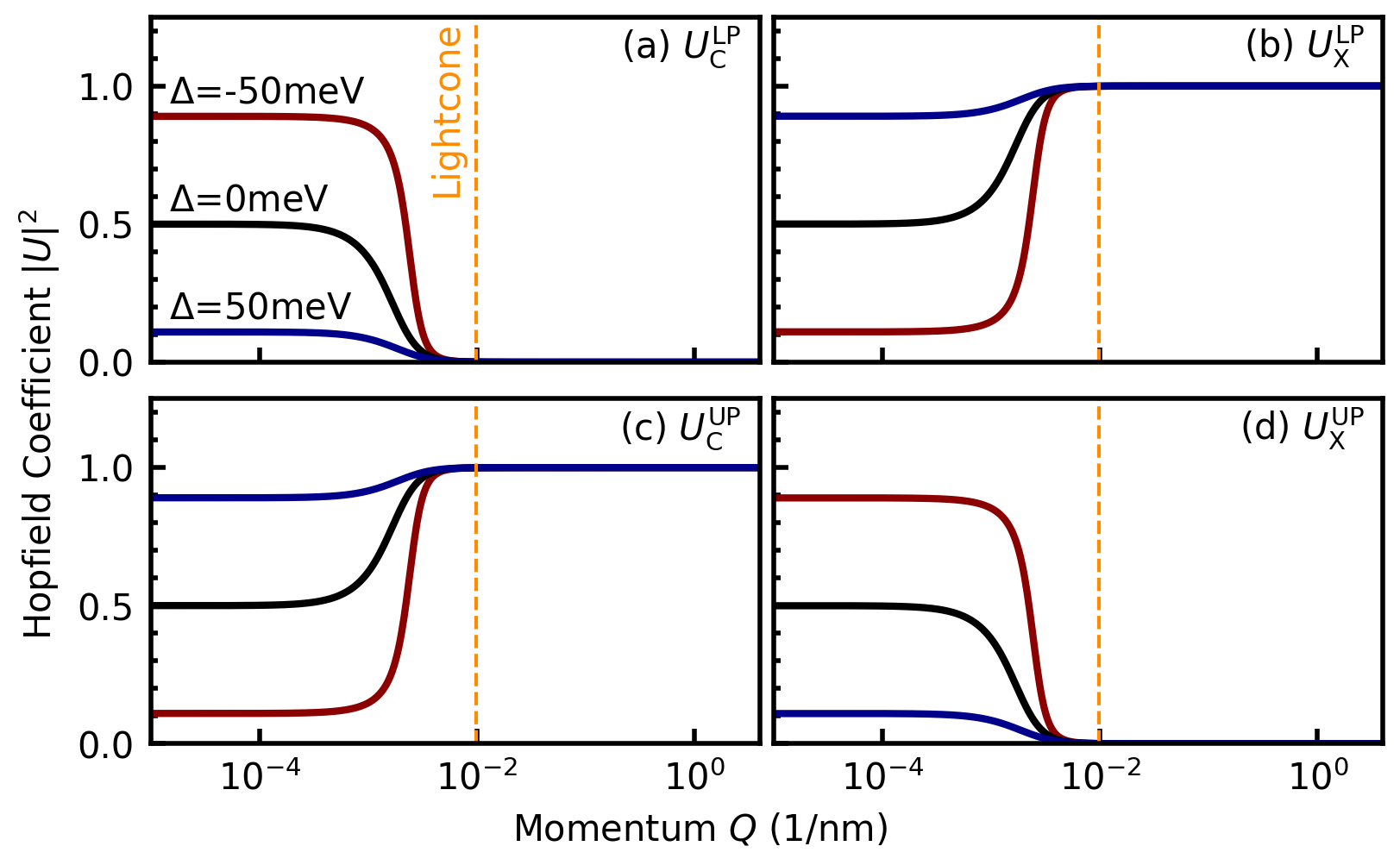}
     \caption{(a) Photonic and (b) excitonic Hopfield coefficients of the lower polariton (LP) in monolayer MoS$_2$ vs. in-plane momentum, $Q$, at different detunings. The vertical orange dashed line denotes the edge of the lightcone. (c) and (d) The same quantities for the upper polariton (UP).}
     \label{fig:S1}
 \end{figure}
\begin{figure}[!t]
     \centering
     \includegraphics[width=.75\textwidth]{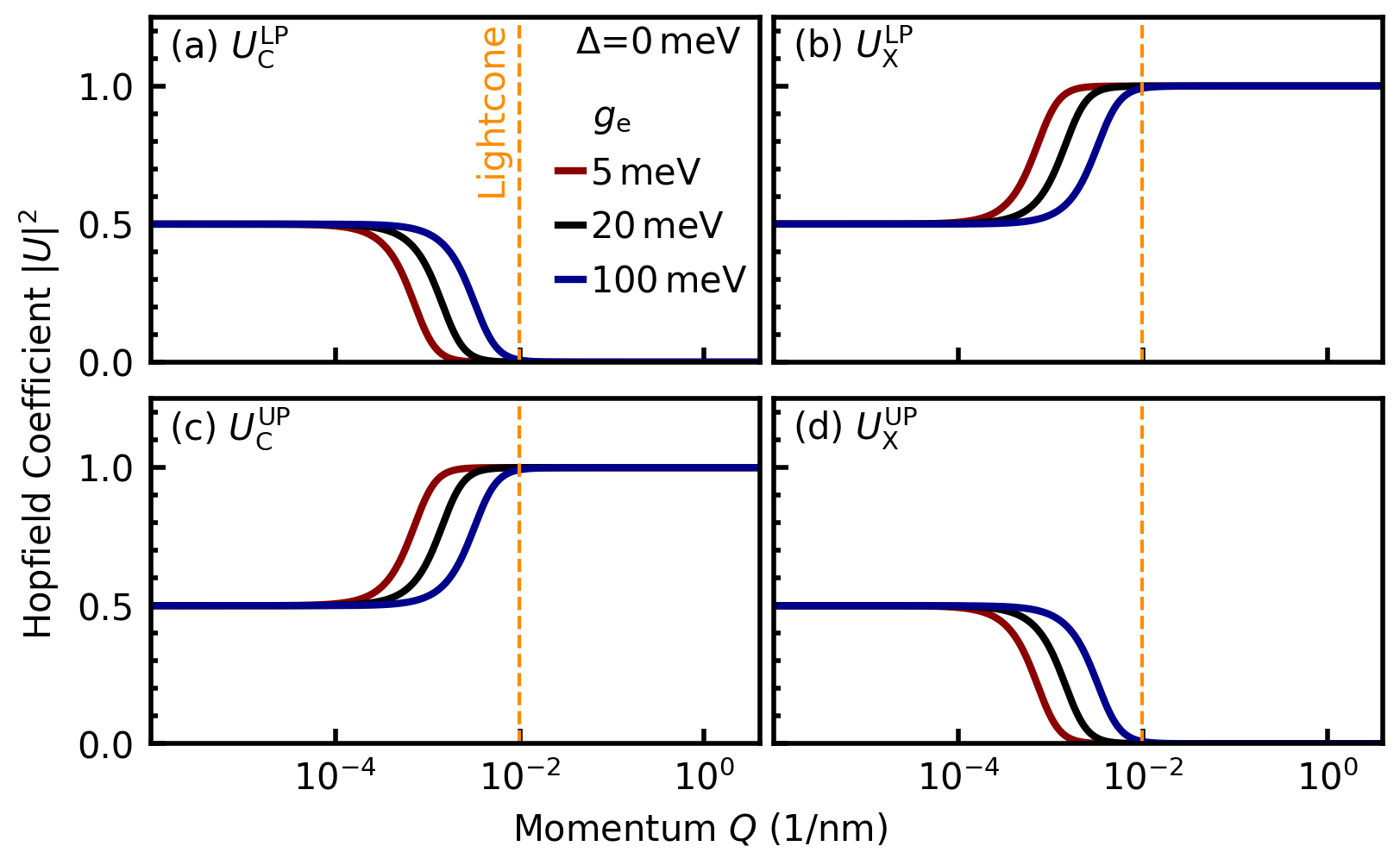}
     \caption{(a) Photonic and (b) excitonic Hopfield coefficients of the lower polariton in monolayer MoS$_2$ vs. in-plane momentum, $Q$, at different coupling strengths, $g_\text{e}$, at zero detuning. The vertical orange dashed line denotes the edge of the lightcone. (c) and (d) The same quantities for the upper polariton.}
     \label{fig:S2}
 \end{figure}

To investigate the strong coupling regime for the monolayer case, we employ the Hopfield approach and diagonalize the following Hamiltonian:
\begin{align}
    \hat H = \sum_{\bm{\bm{Q}}}E_{\bm{Q}}^{(X)} \hat{X}_{\bm{Q}}^\dagger \hat{X}_{\bm{Q}}+E_{\bm{Q}}^{(C)}\hat{c}_{\bm{Q}}^\dagger \hat{c}_{\bm{Q}}+g^{(X)} \hat{X}_{\bm{Q}}^\dagger\hat{c}_{\bm{Q}}+ \text{h.c.}\;\text{,}\label{eq:hopfield_mono}
\end{align}
where $\hat{c}_{\bm{Q}}^{(\dagger)}$ are the photonic annihilation (creation) operators and $E_{\bm{Q}}^{(C)}$ is the cavity photon energy. Note that we only consider the 1s exciton for the monolayer and have therefore omitted the exciton index. The resulting polariton eigenvalues take the form
\begin{align}
    E^{(P)}_{\nu,\bm{Q}}=\frac{E_{\bm{Q}}^{(X)}+E_{\bm{Q}}^{(C)}}{2}\pm\frac{1}{2}\sqrt{\Delta^2+4(g^{(X)})^2}\;\text{,}\label{eq:level}
\end{align}
where we defined the detuning $\Delta=E_{\bm{Q}}^{(X)}-E_{\bm{Q}}^{(C)}$ and the - and + solutions refer to the lower- (LP) and upper (UP) polariton branch, respectively. From the square-root term in Eq.~(\ref{eq:level}), it is evident that the energy splitting between the UP and LP scales linearly for large detuning with respect to the exciton-photon coupling, $\Delta \gg 2g^{(X)}$. Conversely, an expansion for small detunings relative to the coupling strength reveals that the splitting scales quadratically around $\Delta = 0$. Because $\Delta$ and $g_\text{e}$ appear equivalently in the square-root term of Eq.~\ref{eq:level}, the same scaling behavior holds for small and large $g_\text{e}$ as well. The analytical form of the Hopfield coefficients are given by \cite{hopfield1958theory}
\begin{subequations}
\begin{align}
    U_X^\nu(\bm{Q})&=\mp\sqrt{\frac{E^{(P)}_{\nu,\bm{Q}}-E^{(C)}_{\bm{Q}}}{2E^{(P)}_{\nu,\bm{Q}}-E^{(C)}_{\bm{Q}}-E^{(X)}_{\bm{Q}}}}\;\text{,}\\
    U_C^\nu(\bm{Q})&=\sqrt{\frac{E^{(P)}_{\nu,\bm{Q}}-E^{(X)}_{\bm{Q}}}{2E^{(P)}_{\nu,\bm{Q}}-E^{(C)}_{\bm{Q}}-E^{(X)}_{\bm{Q}}}}\;\text{.}
\end{align}
Notably, the excitonic and photonic coefficients have opposite signs for the LP branch and the same sign for the UP branch. Figures~\ref{fig:S1} and \ref{fig:S2} show the Hopfield coefficients as a function of detuning and coupling strength.
\end{subequations}

For the bilayer case, multiple hybrid excitons need to be included in the Hamiltonian:
\begin{align}
\begin{split}
    \hat{H}=&\sum_{\eta,\bm{Q}}E_{\eta,\bm{Q}}^{(Y)}\hat{Y}_{\eta,\bm{Q}}^\dagger \hat{Y}_{\eta,\bm{Q}}+\sum_{\bm{Q}} E_{\bm{Q}}^{(C)}\hat{c}_{\bm{Q}}^\dagger \hat{c}_{\bm{Q}}
    +\sum_{\bm{Q},\eta}g^{(Y)}_\eta \hat{Y}_{\eta,\bm{Q}}^\dagger\hat{c}_{\bm{Q}}+\text{h.c.}\;\text{.}
\end{split}
\end{align}
After applying a Hopfield transformation, we obtain the hybrid exciton polariton energies as a function of the applied electric field. Alternatively, one can diagonalize the full Hamiltonian in the bare exciton basis, including the tunneling term, to move directly to the hybrid exciton polariton basis. This transformation is described by the \emph{generalized Hopfield coefficients} \cite{konig2023interlayer}, which combine the mixing- and Hopfield coefficients: $\tilde U_L^\nu(\bm{Q})=\sum_\eta C_L^\eta(\bm{Q}) U_\eta^{\nu}(\bm{Q}) $ and $\tilde U_C^\nu(\bm{Q})=U_C^{\nu}(\bm{Q}) $. In other words, bilayer exciton polaritons can be described in terms of their intra- and interlayer excitonic character.
\subsection{Fermionic phase-space-filling}
\label{subsec:PSF}
In our investigation of the nonlinear interactions between polaritons due to phase-space-filling (PSF) and the associated energy shifts, we start from the following carrier-light Hamiltonian
\begin{align}
\begin{split}
    \hat{H}_{c-e}=&\sum_{\bm{Q},\bm{q},l,l',s}g_\text{e}^{ll'}\hat{a}^\dagger_{c,l,s,\bm{Q}+\bm{q}}\hat{a}_{v,l',s,\bm{Q}} \hat{c}_{\bm{q}}+(g_\text{e}^{ll'})^{*}\hat{c}_{\bm{q}}^\dagger \hat{a}^\dagger_{v,l',s,\bm{Q}}\hat{a}_{c,l,s,\bm{Q}+\bm{q}}\;\text{,}
\end{split}
\end{align}
where $\hat{a}^{(\dagger)}_{\lambda,l,s,\bm{k}}$ is the annihilation (creation) operator of an electron in band $\lambda$, layer $l$ with spin $s$ and momentum $\bm{k}$, while $\hat{c}_{\bm{q}}^{(\dagger)}$ is the annihilation (creation) operator of a photon with momentum $\bm{q}$. The first term describes the absorption of a photon and resulting transition of an electron from the valence to the conduction band, while the second term describes the inverse process; both processes are governed by the electron-photon matrix element $g_\text{e}^{ll'}$. We enforce spin conservation, as photons cannot induce a spin flip. In this work, we assume $g_\text{e}^{ll'} = g_\text{e} \delta_{ll'}$, since the spatial separation of electrons and holes in different layers strongly suppresses their coupling to the cavity photon, making interlayer electron–photon interactions negligible compared to the intralayer ones \cite{jiang2021interlayer}. We then derive the equation of motion for the microscopic polarization $\hat{p}_{\bm{k}+\bm{Q}',\bm{k}}^{\dagger l_e,l_h,s}=\hat{a}_{c,l_e,s,\bm{k}+\bm{Q}'}^\dagger \hat{a}_{v,l_h,s,\bm{k}}$ via the Heisenberg's equation of motion yielding
\begin{align}
\begin{split}
    i\hbar\dv{t} \langle \hat{p}_{\bm{k}+\bm{Q}',\bm{k}}^{\dagger l_e,l_h,s}\rangle\big|_{\hat{H}_{c-e}}=&-(g_\text{e}^{l_el_h})^* \langle \hat{c}_{\bm{Q}'}^\dagger\rangle\\
    &+\sum_{\bm{q},l}(g_\text{e}^{ll_h})^*\langle \hat{c}_{\bm{q}}^\dagger \hat{a}_{c,l_e,s,\bm{k}+\bm{Q}'}^\dagger \hat{a}_{c,l,s,\bm{k}+\bm{q}}\rangle\\
    &+(g_\text{e}^{l_el})^*\langle \hat{c}_{\bm{q}}^\dagger \hat{a}_{v,l_h,s,\bm{k}_1}\hat{a}_{v,l,s,\bm{k}_1+\bm{Q}-\bm{q}}^\dagger\rangle\;\text{,}\label{eq:psf_eh}
\end{split}
\end{align}
where the first term describes the conventional electron-photon coupling \cite{fitzgerald2022twist,fitzgerald2025polariton}, while the other two terms describe a reduction of this coupling due to the electron and hole densities via Pauli blocking, representing an effective saturation of the coupling.

Following the approach presented in Ref.~\cite{erkensten2023electrically}, we use $\hat{a}_{c,l,s,\bm{k}}^\dagger \hat{a}_{c,l',s,\bm{k}'}\approx\sum_{l'',\bm{k}''}\hat{p}_{\bm{k},\bm{k}''}^{\dagger l,l'',s}\hat{p}_{\bm{k}',\bm{k}'',s}^{ l',l'',s}$ and $
\hat{a}_{v,l,s,\bm{k}} \hat{a}_{v,l',s,\bm{k}'}^\dagger\approx\sum_{l'',\bm{k}''}\hat{p}_{\bm{k}'',\bm{k}}^{\dagger l'',l,s}\hat{p}_{\bm{k}'',\bm{k}'}^{l'',l',s}$ and then transform Eq.~(\ref{eq:psf_eh}) into the exciton basis via $\hat{p}_{\bm{k},\bm{k}'}^{l_e,l_h,s}=\Psi_{\beta\bm{k}+\alpha\bm{k}'}^{(l_e,l_h,s)}\hat{X}_{(l_e,l_h,s),\bm{k}-\bm{k}'}$ and $
\hat{p}_{\bm{k},\bm{k}'}^{\dagger l_e,l_h,s}=\Psi_{\beta\bm{k}+\alpha\bm{k}'}^{*(l_e,l_h,s)}\hat{X}^\dagger_{(l_e,l_h,s),\bm{k}-\bm{k}'}$, where $\alpha=\frac{m_e}{m_h+m_e}$ and $\beta=\frac{m_h}{m_h+m_e}$, which we approximate to be independent of the spin and layer configuration as the effective masses at the K points vary minimally between bands \cite{laturia2018dielectric}. Then Eq.~(\ref{eq:psf_eh}) becomes
\begin{subequations}
\begin{align}
\begin{split}
    i\hbar\dv{t}\left\langle \hat{X}_{L,\bm{Q}'}^{\dagger}\right\rangle\big|_{\hat{H}_{c-e}}=&-(g_L^{(X)})^*\langle \hat{c}_{\bm{Q}'}^\dagger\rangle+\sum_{\bm{q},\bm{Q},L'}g_\text{e}^*R_{\bm{Q}',\bm{Q},\bm{q}}^{LL'}\left\langle \hat{c}_{\bm{q}}^\dagger \hat{X}_{L,\bm{Q}+\bm{Q}'-\bm{q}}^{\dagger} \hat{X}_{L',\bm{Q}}\right\rangle\;\text{,}\label{eq:psf_bilayer_exc}
\end{split}
\end{align}
where
\begin{align}
    R_{\bm{Q}',\bm{Q},\bm{q}}^{LL'}&=\delta_{l_el_h}\delta_{ss'}(\delta_{l_e,l_e'}R_{e,\bm{Q}',\bm{Q},\bm{q})}^{l_el_h's}+\delta_{l_h,l_h'}R_{h,\bm{Q}',\bm{Q},\bm{q}}^{l_e'l_hs})\;\text{,}\\
    R_{e,\bm{Q}',\bm{Q},\bm{q}}^{ll's}&=\sum_{\bm{k}}\Psi_{\bm{k}}^{(l,l,s)}\Psi_{\bm{k}+\alpha(\bm{q}-\bm{Q})}^{*(l,l',s)}\Psi_{\bm{k}+\bm{q}-\beta \bm{Q}'-\alpha \bm{Q}}^{(l,l',s)}\;\text{,}\\
    R_{h,\bm{Q}',\bm{Q},\bm{q}}^{l'ls}&=\sum_{\bm{k}}\Psi_{\bm{k}}^{(l,l,s)}\Psi_{\bm{k}+\beta(\bm{Q}-\bm{q})}^{*(l',l,s)}\Psi_{\bm{k}-\bm{q}+\beta \bm{Q}+\alpha \bm{Q}'}^{(l',l,s)}\;\text{.}
\end{align}
\end{subequations}
In the case of the monolayer, we include only the 1s exciton. Therefore, we omit spin/layer indices and transform Eq.~(\ref{eq:psf_bilayer_exc}) into the polariton basis with $ \hat{X}_{\bm{Q}}^{\dagger}=\sum_\nu U_X^{\nu}(\bm{Q}) \hat{P}_{\nu,\bm{Q}}^{\dagger}$ and $\hat{c}_{\bm{Q}}^\dagger=\sum_\nu U_C^{\nu}(\bm{Q}) \hat{P}_{\nu,\bm{Q}}^\dagger$. Then Eq.~(\ref{eq:psf_bilayer_exc}) becomes
\begin{subequations}
\begin{align}
\begin{split}
    i\hbar \dv{t}\left\langle \hat{P}_{\nu,\bm{Q}'}^\dagger\right\rangle\big|_{\hat{H}_{c-e}}=&-\left\langle \hat{P}_{\nu,\bm{Q}'}^\dagger\right\rangle \Delta E_{\mathrm{PSF}}^\nu(\bm{Q}')\;\text{,} \label{eq:psf_pol_mono}   
\end{split}
\end{align}
where we obtain the energy renormalization due to fermionic PSF (saturation interaction) as
\begin{align}
\begin{split}
    \Delta E_{\mathrm{PSF}}^\nu(\bm{Q}')=&-g_\text{e}^*\sum_{\bm{Q},\nu'}\Bigl[R_{1,\bm{Q}',\bm{Q}}U_C^\nu(\bm{Q}')U_X^{*\nu}(\bm{Q}')\left|U_X^{\nu'}(\bm{Q})\right|^2\\
    &+R_{2,\bm{Q}',\bm{Q}}U_C^{\nu'}(\bm{Q})U_X^{*\nu'}(\bm{Q})|U_X^\nu(\bm{Q}')|^2\Bigr]N_{\nu',\bm{Q}}\;\text{,}\label{eq:shift_psf_monolayer}
\end{split}
\end{align}
\end{subequations}
with $R_{1,\bm{Q}',\bm{Q}}=R_{\bm{Q}',\bm{Q},\bm{Q}'}$, $R_{2,\bm{Q}',\bm{Q}}=R_{\bm{Q}',\bm{Q},\bm{Q}}$ and introducing the polariton occupation $N_{\nu',\bm{Q}}=\left\langle \hat{P}_{\nu',\bm{Q}}^\dagger \hat{P}_{\nu',\bm{Q}}\right\rangle$. In order to derive the equation above, we have performed the cluster expansion within the random-phase approximation. To capture the density dependence, we restrict the expectation values to diagonal terms, where the polariton indices coincide. Our approach is therefore perturbative and does not account for interaction-induced modifications of the polariton character. While the first term of Eq.~(\ref{eq:shift_psf_monolayer}) describes how polaritons inside the lightcone are affected by excitons in- and outside the lightcone, the latter describes how excitons are affected by polaritons. In particular, the second term also captures the modification of the excitonic reservoir due to a large polariton occupation inside the lightcone. In the case of $\bm{Q}'=0$, Eq.~(\ref{eq:shift_psf_monolayer}) reduces to the second term of equation (1) of the main text. Importantly, the first term of Eq.~(\ref{eq:psf_bilayer_exc}) describes the strong coupling between the exciton and photon and is reflected as the last term in Eq. (\ref{eq:hopfield_mono}). Therefore, when performing the Hopfield transformation, this term gets absorbed into the diagonal part of the polariton Hamiltonian and does not need to be accounted for in Eq.~ (\ref{eq:psf_pol_mono}). 

Focusing on $\bm{Q}'=0$, the UP can exhibit a blueshift due to the saturation interaction, but only at cryogenic temperatures and large red detunings, in contrast to the usual expected redshift, as shown in Fig.~2(e) of the main text. We emphasize that, at any realistic temperature, only the LP is significantly occupied. Thus, we can rewrite the energy-shift of the UP in Eq.~\ref{eq:shift_psf_monolayer} as 
\begin{align}
\begin{split}
    \Delta E_{\mathrm{PSF}}^\text{UP}(0)=&-g_\text{e}\sum_{\bm{Q}}\left[R_{1,0,\bm{Q}} U_C^\text{UP}(0)U_X^{*\text{UP}}(0)\left|U_X^\text{LP}(\bm{Q})\right|^2\right.\\
    &\left.+R_{2,0,\bm{Q}} U_C^\text{LP}(\bm{Q})U_X^{*\text{LP}}(\bm{Q})\left|U_X^\text{UP}(0)\right|^2 \right]N_{\text{LP},\bm{Q}}\;\text{.}
\end{split}\label{eq:shift_psf_mono_q0}
\end{align}
For high temperatures, the exciton reservoir is populated and therefore the photonic character of the majority of the occupied LP states is negligible, making the first term of Eq. (\ref{eq:shift_psf_mono_q0}) dominate. As discussed in section~\ref{subsec:Hopfield}, the excitonic and photonic Hopfield coefficients of the UP have the same sign, which makes the first term negative and thus leads to redshift. On the other hand, for a red-detuned system at low temperatures, the occupied LP states become very photonic and much less excitonic. While the first term of Eq.~\ref{eq:shift_psf_mono_q0} scales quadratically with the excitonic Hopfield coefficient of LP, the second term scales only linearly, making this the dominant term. If the detuning is not too large, the second term will become positive due to the excitonic and photonic Hopfield coefficients of the LP having opposite signs, leading to a blueshift of the UP in this low temperature regime.

Turning to the bilayer case, we return to Eq.~(\ref{eq:psf_bilayer_exc}) and transform it directly to the hybrid exciton polariton picture via the transformation $\hat{X}_{L,\bm{Q}}^\dagger=\sum_{\nu}\hat{P}_{\nu,\bm{Q}}^\dagger \tilde U_{L}^{\nu}(\bm{Q})$ and $\hat{c}_{\bm{Q}}^\dagger=\sum_{\nu}\hat{P}_{\nu,\bm{Q}}^\dagger \tilde U_{C}^{\nu}(\bm{Q})$ that was discussed in section~\ref{subsec:Wannier_hyb}, in analogy to the monolayer case. We obtain
\begin{subequations}
\begin{align}
    i\hbar\dv{t}\left\langle \hat{P}_{\nu,\bm{Q}'}^{\dagger}\right\rangle\big|_{\hat{H}_{c-e}}=&-\left\langle \hat{P}_{\nu,\bm{Q}'}^{\dagger}\right\rangle\Delta E_{\mathrm{PSF}}^\nu(\bm{Q}')\;\text{,}
\end{align}
where
\begin{align}
    \Delta E_{\mathrm{PSF}}^\nu(\bm{Q}')=-g_\text{e}\sum_{\bm{Q},\nu'}\left(\mathcal{R}^{\nu\nu'}_{1,\bm{Q}'\bm{Q}}+\mathcal{R}^{\nu\nu'}_{2,\bm{Q}'\bm{Q}}\right) N_{\nu',\bm{Q}}\;\text{,}\label{eq:shift_psf_bilayer}
\end{align}
and 
\begin{align}
    \mathcal{R}^{\nu\nu'}_{1,\bm{Q}'\bm{Q}}&=\sum_{LL'}R^{LL'}_{1,\bm{Q}'\bm{Q}}\tilde U_{L}^{*\nu}(0)\tilde U_{C}^{\nu}(0)\tilde U_{L'}^{\nu'}(\bm{Q})\tilde U_{L}^{*\nu'}(\bm{Q})\;\text{,}\\
    \mathcal{R}^{\nu\nu'}_{2,\bm{Q}'\bm{Q}}&=\sum_{LL'}R^{LL'}_{2,\bm{Q}'\bm{Q}}\tilde U_{L}^{*\nu}(0)\tilde U_{C}^{\nu'}(0)\tilde U_{L'}^{\nu}(\bm{Q})\tilde U_{L}^{*\nu'}(\bm{Q})\;\text{,}
\end{align}
with $R^{LL'}_{1,\bm{Q}'\bm{Q}}=R_{\bm{Q',\bm{Q},\bm{Q'}}}^{LL'}$ and $R^{LL'}_{2,\bm{Q}'\bm{Q}}=R_{\bm{Q',\bm{Q},\bm{Q}}}$. Note that Eq.~(\ref{eq:shift_psf_bilayer}) coincides with Eq.~(\ref{eq:shift_psf_monolayer}) when including only a single exciton, where the mixing coefficients become unity and the generalized and regular Hopfield coefficients become the same. Furthermore, for $\bm{Q}'=0$, Eq.~(\ref{eq:shift_psf_bilayer}) reduces to the second term in Eq.~(2) of the main text.

\end{subequations}
\subsection{Fermionic exchange interaction}
To include the fermionic exchange interaction stemming from the indirect term of the Coulomb interaction which is the dominant contribution to the exciton-exciton interaction in TMD monolayers \cite{erkensten2021exciton, Shahnazaryan2017exciton}, we start from the corresponding part of the equation of motion in the exciton basis, taken from Ref.~\cite{erkensten2023electrically} while omitting any valley or layer index:
\begin{align}
    i\hbar\dv{t}\left\langle \hat{X}_{\bm{Q}'}^{\dagger}\right\rangle\big|_\text{exch.}=-&\sum_{\bm{q},\bm{Q}}W_{(\bm{Q}',\bm{Q},\bm{q})}\left\langle \hat{X}_{\bm{Q}'+\bm{q}}^\dagger \hat{X}_{\bm{Q}-\bm{q}}^\dagger \hat{X}_{\bm{Q}}\right\rangle\;\text{,}\label{eq:exc_ex_mono}
\end{align} 
with the form of the matrix elements $W_{\bm{Q}',\bm{Q},\bm{q}}$ given in Ref.~\cite{erkensten2023electrically}. We now transform Eq.~(\ref{eq:exc_ex_mono}) into the polariton basis using the same transformation as in section~\ref{subsec:Wannier_hyb} and obtain 
\begin{align}
    i\hbar \dv{t}\left\langle \hat{P}_{\nu,\bm{Q}'}^{\dagger}\right\rangle\big|_\text{exch.}=&-\left\langle \hat{P}_{\nu,\bm{Q}'}^{\dagger}\right\rangle\Delta E_\text{exch.}^\nu(\bm{Q}')\;\text{,}
\end{align}
where
\begin{align}
    \Delta E_\text{exch.}^\nu(\bm{Q}')&=\sum_{\nu',\bm{Q}'}|U_X^\nu(\bm{Q}')|^2|U_X^{\nu'}(\bm{Q})|^2W_{\bm{Q}'\bm{Q}}N_{\nu',\bm{Q}}^{(P)}\;\text{,}
\end{align}
with $W_{\bm{Q}'\bm{Q}}=W_{\bm{Q}',\bm{Q},0}+W_{\bm{Q}',\bm{Q},\bm{Q}-\bm{Q}'}$. For simplicity, we consider the long-wavelength limit of the exchange interaction, $W_{\bm{Q}'\bm{Q}}\approx{W_{0, 0}}\equiv W$, as it typically decays over a much longer momentum scale than the exciton population \cite{tassone1999exciton, erkensten2022microscopic}. Then the energy shift at $\bm{Q}=0$ becomes
\begin{align}
    \Delta E_\text{exch.}^\nu(0)&\approx W\sum_{\nu',\bm{Q}'}|U_X^\nu(\bm{Q}')|^2|U_X^{\nu'}(\bm{Q})|^2N_{\nu',\bm{Q}}^{(P)}\;\text{,}\label{eq:shift_exch_mono}
\end{align}
with $W=2\sum_{\bm{k},\bm{k}'} V_{|\bm{k}-\bm{k}'|}\left(\Psi_{\bm{k}'}-\Psi_{\bm{k}}\right)\Psi_{\bm{k}}^*|\Psi_{\bm{k}'}|^2$. Eq. (\ref{eq:shift_exch_mono}) corresponds to the first term of Eq.~(1) in the main text. Using the wave function obtained by solving the Wannier equation (\ref{eq:wannier_rel}), we find $W\approx 0.9$\,eV\,nm$^2$, which is in line with other studies \cite{Shahnazaryan2017exciton,erkensten2021exciton,mantsevich2024viscous}. The exchange interaction of 1s excitons can be shown to be directly proportional to $E_b^{\mathrm{1s}}a_B^2$, where $a_B$ is the exciton Bohr radius and is therefore only weakly dependent on dielectric environment \cite{Shahnazaryan2017exciton, ciuti1998role}. 

\subsection{Dipole-dipole interaction}
For the dipole-dipole part of the exciton-exciton interaction, which constitutes the dominant contribution in TMD bilayers, we take the corresponding part of the equation of motion in the exciton picture from Ref.~\cite{erkensten2023electrically}:
\begin{align}
    i\hbar\dv{t}\left\langle \hat{X}_{L,\bm{Q}'}^{\dagger}\right\rangle\big|_\text{dip.}=-&\sum_{\bm{q},\bm{Q},L'}{D}_{L'L,\bm{q}}\left\langle \hat{X}_{L\bm{Q}'+\bm{q}}^\dagger \hat{X}_{L',\bm{Q}-\bm{q}}^\dagger \hat{X}_{L',\bm{Q}}\right\rangle\;\text{,}\label{eq:dip_ex_bi}
\end{align}
with the form of the matrix elements ${D}_{L'L,\bm{q}}$ given in Ref.~\cite{erkensten2023electrically}. Transforming into the polariton basis gives
\begin{subequations}
\begin{align}
    i\hbar \dv{t}\left\langle \hat{P}_{\nu,\bm{Q}'}^{\dagger}\right\rangle\big|_\text{dip.}=&-\left\langle \hat{P}_{\nu,\bm{Q}'}^{\dagger}\right\rangle\Delta E_\text{dip.}^\nu(\bm{Q}')\;\text{,}
\end{align}
where
\begin{align}
    \Delta E_\text{dip.}^\nu(\bm{Q}')&=\sum_{\bm{Q},\nu'}\left(\mathcal{D}^{\nu\nu'}_{1,\bm{Q}'\bm{Q}}+\mathcal{D}^{\nu\nu'}_{2,\bm{Q}'\bm{Q}}\right)N_{\nu',\bm{Q}}^{(P)}\;\text{,}\label{eq:shift_dip_bilayer}
\end{align}
and 
\begin{align}
    \mathcal{D}^{\nu\nu'}_{1,\bm{Q}'\bm{Q}}&=\sum_{L,L'}D_{L'L,0} |\tilde U_{L}^\nu(\bm{Q}')|^2|\tilde U_{L'}^{\nu'}(\bm{Q})|^2\;\text{,}\\
    \mathcal{D}^{\nu\nu'}_{2,\bm{Q}'\bm{Q}}&=\sum_{L,L'}D_{LL',\bm{Q}-\bm{Q'}}\tilde U_{L}^{*\nu}(\bm{Q}')\tilde U_{L'}^{\nu}(\bm{Q}')\tilde U_{L'}^{\nu'}(\bm{Q})\tilde U_{L}^{*\nu'}(\bm{Q})\;\text{.}
\end{align}
\end{subequations}
In the case of $\bm{Q}'=0$, Eq.~(\ref{eq:shift_dip_bilayer}) reduces to the first term of Eq.~(2) in the main text.

\subsection{Polaritonic Elliott formula}
To calculate the absorption of polaritons we use the polaritonic Elliott formula \cite{fitzgerald2022twist}
\begin{align}
    A(\hbar\omega,\bm{Q})&=\frac{2\kappa\left(\Re(\Pi(\hbar\omega)-\Gamma|\Pi(\hbar\omega)|^2\right)}{|1+(\kappa-\Gamma)\Pi(\hbar\omega)|^2}\;\text{,}\\
    \Pi(\hbar\omega)&=\sum_\nu\frac{|U_C^\nu(\bm{Q}')|^2}{i(E_{\nu,\bm{Q}'}^{(P)}+\Delta E_\nu(\bm{Q}')-\hbar\omega)+\Gamma}\;\text{,}\label{eq:pol_elliott1}
\end{align}
where $\Gamma$ and $\kappa=\frac{c T_m}{2L_c}$ are the exciton dephasing and cavity photon radiative decay rate, respectively. Here, $T_m$ is the cavity transmission coefficient and $L_c$ the cavity length. For illustrative purposes, a temperature-independent excitonic linewidth of $\Gamma=1\,\mathrm{meV}$ is assumed in this study. In realistic systems, the linewidth typically exhibits a strong temperature dependence, however, this would not qualitatively alter the conclusions. We can approximate Eq.~(\ref{eq:pol_elliott1}) by assuming energetically well-spaced polaritons:
\begin{align}
    A(\hbar\omega,\bm{Q})&=\sum_{\nu}\frac{2\tilde\kappa_{\nu,\bm{Q}'}\tilde\Gamma_{\nu,\bm{Q}'}}{(E_{\nu,\bm{Q}'}^{(P)}+\Delta E_\nu(\bm{Q}')-\hbar\omega)^2+(\tilde\kappa_{\nu,\bm{Q}'}+\tilde\Gamma_{\nu,\bm{Q}'})^2}\;\text{,}\label{eq:pol_elliott2}
\end{align}
where $\tilde\kappa_{\nu,\bm{Q}'}=|U_C^\nu(\bm{Q}')|^2\kappa$ and $\tilde\Gamma_{\nu,\bm{Q}'}=(1-|U_C^\nu(\bm{Q}')|^2)\Gamma$ describe the photonic and material-based decay channels of the polariton, respectively. Due to symmetry reasons, the resonant absorption is limited to the maximum value of 0.5 \cite{piper2014}, which is reached when $\tilde\kappa_{\nu,\bm{Q}'}=\tilde\Gamma_{\nu,\bm{Q}'}$ is fulfilled. This is called the critical coupling condition \cite{fitzgerald2022twist,ferreira2024revealing} and can be rewritten as a condition on the photonic Hopfield coefficient of the corresponding polariton branch $|U_C^\nu(\bm{Q})|^2=\frac{\Gamma}{\Gamma+\kappa}$. Therefore, when excitonic and photonic losses are similar in magnitude, the polariton must retain a substantial balance of photonic and excitonic components to reach maximum absorption.
 
 \section{Polaritons in a MoS$_2$ homobilayer under an electric field}
 \subsection{Polariton energy landscape}
 \begin{figure}[!t]
     \centering
     \includegraphics[width=1.\textwidth]{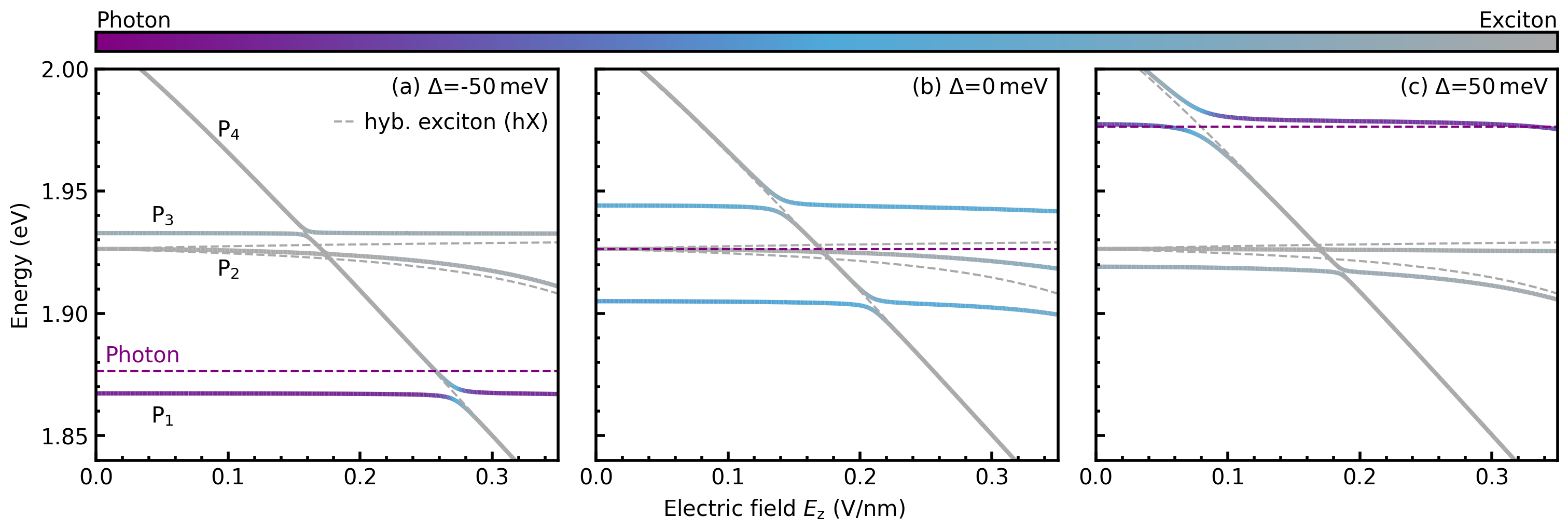}
     \caption{(a) Polariton energy landscape in a 2H-stacked MoS$_2$ homobilayer as a function of an applied out-of-plane electric field $E_\mathrm{z}$ for a detuning of $\Delta=-50\,\mathrm{meV}$. The solid lines denote the energies of the four lowest-energy polariton branches, $P_1$ to $P_4$, with the color gradient showing the polariton's excitonic/photonic character. Gray and purple dashed lines indicate the hybrid exciton branches and the cavity photon energy, respectively. (b) and (c) The same as (a) at $\Delta=0\,\mathrm{meV}$ and $+50$\,meV, respectively. }
     \label{fig:S3}
 \end{figure}
 
 \begin{figure}[!t]
     \centering
     \includegraphics[width=.75\textwidth]{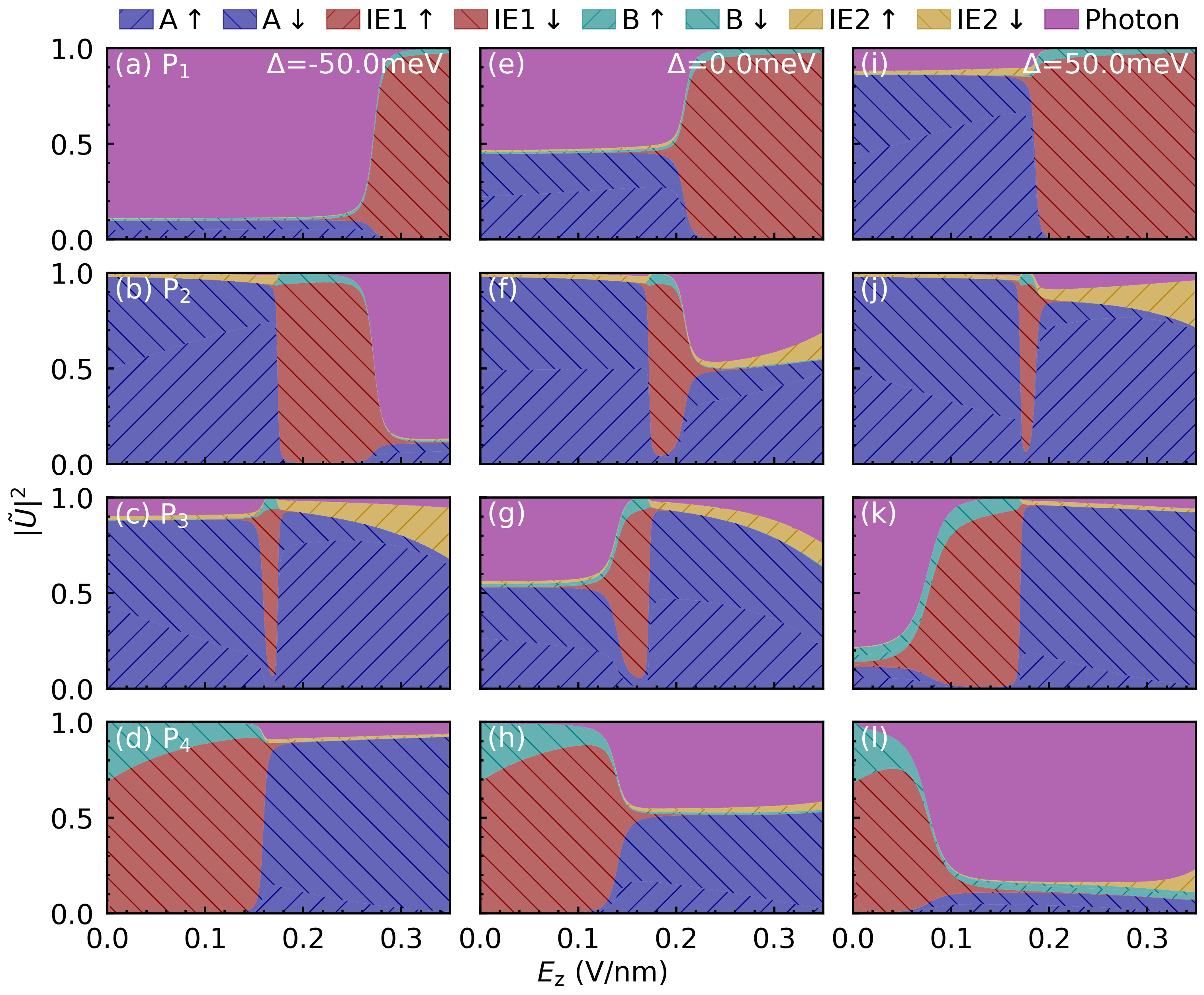}
     \caption{Generalized Hopfield coefficients $|\tilde U|^2$ as a function of electric field $E_\text{z}$. Panels (a)-(l) show the composition of the polariton branches $P_1$-$P_4$, arranged by rows (top to bottom) and detunings $\Delta$=-50, 0, and $50\,\mathrm{meV}$ by columns (left to right).}
     \label{fig:S4}
 \end{figure}
 
 While the polariton landscape under an out-of-plane electric field for a red-detuned cavity has been discussed in the main manuscript (Figs.~3(a) and \ref{fig:S3}(a)), we focus here on the other detuning regimes. For small electric fields, $E_\text{z}\leq0.1\,\mathrm{V/nm}$, we have a situation similar to the case of a single photon coupling to a single exciton (cf. Fig. 2(a) for normal incidence), since the other hybrid excitons lie at significantly higher energies and therefore can not couple strongly to the photon, as shown in Fig.~\ref{fig:S3}(b). In practice, the degeneracy of the intralayer-like hybrid excitons leads to an additional dark, flat middle polariton branch that is purely excitonic (P$_2$) \cite{konig2025magneto}, while P$_1$ and P$_3$ take the role of the lower and upper polariton branch, respectively, with a large Rabi splitting of 40\,meV and a very hybridized character, as shown by the light blue gradient. As the electric field is increased, the interlayer-like hybrid exciton is pushed down in energy and crosses the intralayer-like hybrid excitons due to the forbidden electron tunneling. This results in the two small avoided crossings located at $E_\text{z}$=0.12\,V/nm and 0.22\,V/nm. Upon further increasing the electric field, P$_1$ becomes very excitonic while P$_2$ and P$_4$ now take the role of the upper and lower polariton. However, at higher fields the degeneracy of the intralayer-like hybrid excitons is lifted as higher interlayer-like excitons, which can couple via the allowed hole tunneling, move closer in energy. As a result, the middle polariton branch P$_3$ becomes more hybridized as the field increases. For a blue-detuned cavity, the landscape looks similar to the red-detuned case, with only the cavity photon being above the intralayer-like hybrid excitons and therefore the largest avoided crossing appearing above the almost fully excitonic polaritons at a field strength of $0.09\,\mathrm{V/nm}$.
\begin{figure}[!t]
     \centering
     \includegraphics[width=1.\textwidth]{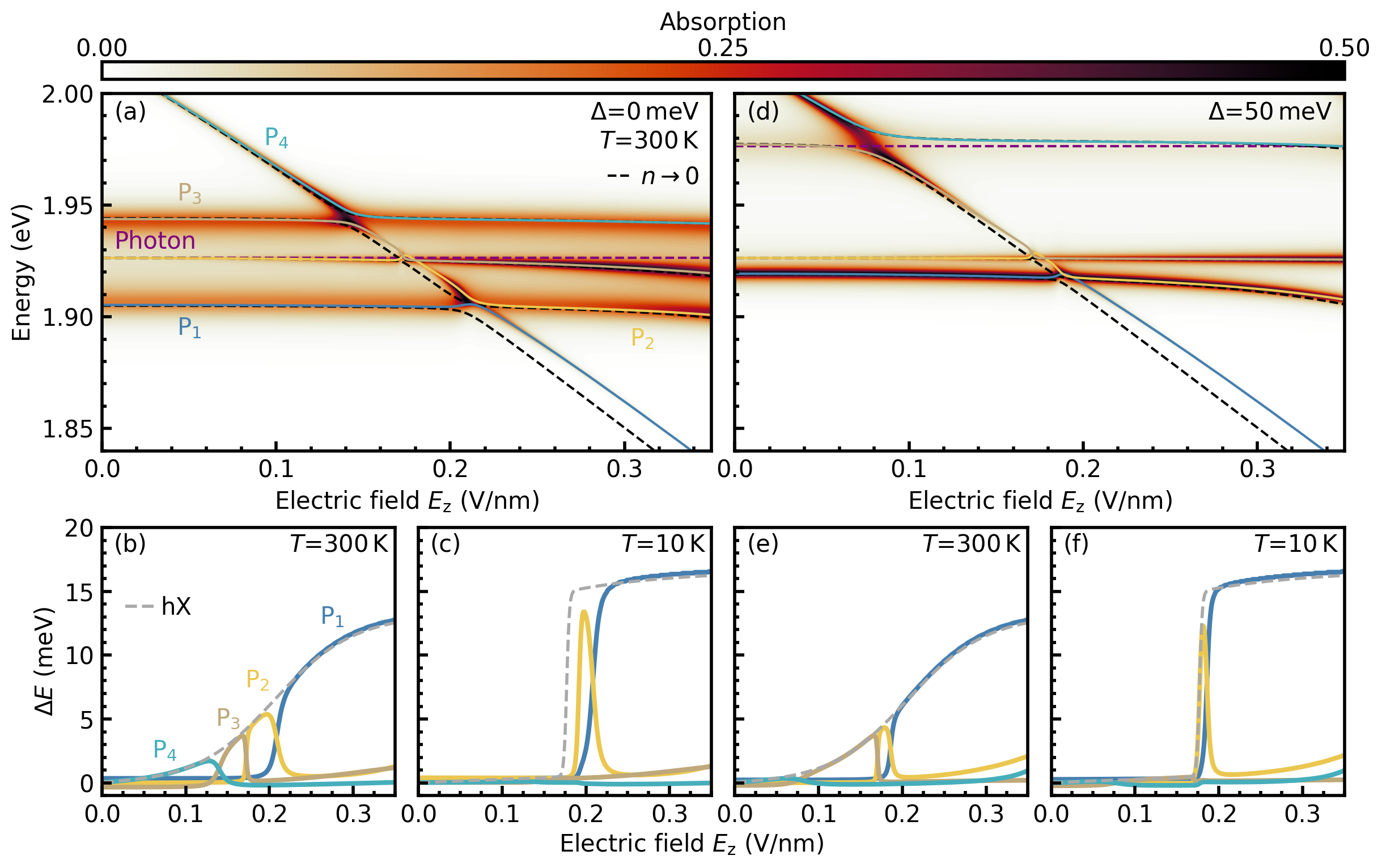}
      \caption{(a) Impact of interaction-induced shifts on the polariton energy landscape in a 2H-stacked MoS$_2$ homobilayer under an out-of-plane electric field $E_\mathrm{z}$. Results are shown for the four lowest polariton branches at a detuning of $\Delta=0\,\mathrm{meV}$, a temperature of $300\,\mathrm{K}$, and a polariton density of $10^{12}/\mathrm{cm^2}$. The colormap shows the polariton absorption. The low-density limit, where the interaction-induced shifts are zero, is denoted by the black dashed lines. (b)-(c) Field-dependent nonlinear energy shifts resolved for different polariton branches in (a) at (b) 300\,K and (c) 10\,K. The gray dashed line shows the shift of the predominantly interlayer-like hybrid exciton. (d)-(f) The same quantities as (a)-(c) for $\Delta=+50\,\mathrm{meV}$.}
     \label{fig:S5}
\end{figure}
Figure~\ref{fig:S4} presents the generalized Hopfield coefficients, $\tilde U_L^\nu$, as a function of the applied electric field for different detuning values. The key observation is that, for any fixed detuning, the polariton branch exhibiting the strongest interlayer character (red area) changes systematically with the increasing electric field. At low fields, this character is predominantly associated with the P$_4$ branch, but as the field strength increases it is successively transferred to P$_3$, then P$_2$, and ultimately to P$_1$. The role of the detuning is mainly to determine the field values at which these transfers occur, as well as to influence which of the involved branches simultaneously carry a significant photonic component.

\subsection{Nonlinear interactions}
\begin{figure}[!t]
     \centering
     \includegraphics[width=.75\textwidth]{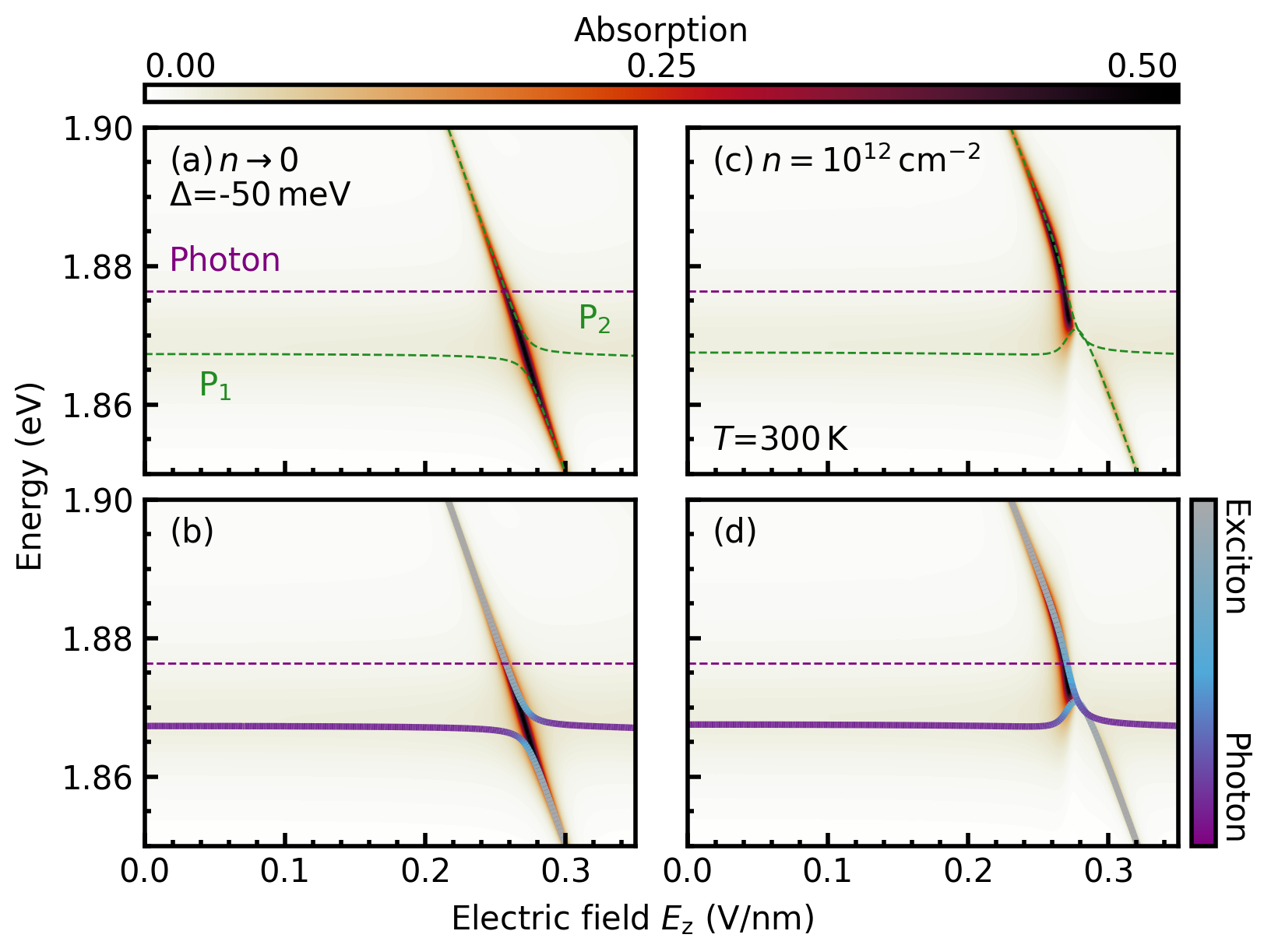}
     \caption{(a) Polariton energy landscape for homobilayer MoS$_2$ around the lowest two polariton branches for a detuning of $\Delta=-50\,\mathrm{meV}$ as a function of electric field $E_\text{z}$ in the low density limit where nonlinear interactions are negligible. The green and purple dashed lines denote the polaritons and photon respectively. The colormap shows the absorption. (b) The same quantities as in (a) now with an added gradient showing the excitonic/photonic contributions to each polariton branch. (c) and (d) The same quantities as (a) and (b) for a density of $n=10^{12}\,\mathrm{cm}^{-2}$ at room temperature}
     \label{fig:S6}
\end{figure}
Figures~\ref{fig:S5}(a)–(c) show the equivalent of Fig.~4 from the main text for zero detuning and Figs.~\ref{fig:S5}(d)–(f) for $\Delta = +50,\mathrm{meV}$. At a given field, the polariton branch with the highest interlayer character (cf.~Fig.~\ref{fig:S4}) shows the same shift as the interlayer-like hybrid exciton at the same density and field strength, as shown in Figs.~\ref{fig:S5}(b) and (c), and \ref{fig:S5}(e) and (f). Compared to a red-detuned cavity, the energy separation between P$_1$ and the lowest-lying hybrid exciton becomes smaller at larger detuning. As a result, interlayer-like polaritons approach occupied states at field strengths similar to those for hybrid excitons in the absence of a cavity, such that at low temperatures, the onset of polariton energy shifts occurs at field strengths comparable to those of the hybrid exciton.

\begin{figure}[!t]
     \centering
     \includegraphics[width=.75\textwidth]{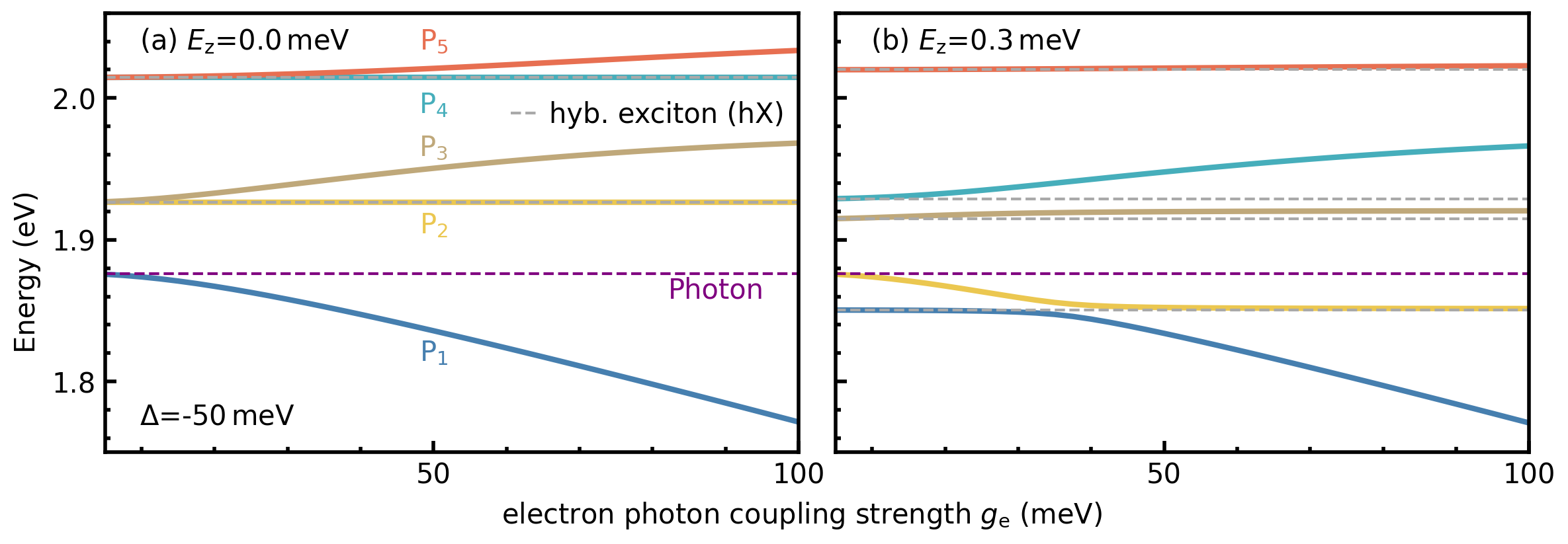}
     \caption{(a) Polariton energy landscape for the lowest 5 branches for 2H-stacked homobilayer MoS$_2$ as a function of the electron photon coupling strength $g_\text{e}$ for a detuning of $\Delta=-50\,\mathrm{meV}$ and zero electric field. The dashed vertical grey and purple lines denote the hybrid exciton and photon energies, respectively. (b) The same quantities as (a) for an electric field strength of $E_\text{z}=0.3\,\mathrm{V/nm}$.}
     \label{fig:S7}
\end{figure}

We now turn to the absorption in the nonlinear regime and compare it to the low-density case. Since the absorption features near the lowest-energy avoided crossings are qualitatively similar for different detunings, we focus here on the red-detuned case. Figure~\ref{fig:S6}(a) shows the polariton landscape in the low-density limit. P$_1$ and P$_2$ form an avoided crossing at 0.27\,V/nm, however this is not visible in absorption due to the condition for strong coupling not being met at these detunings. Absorption along the branches is governed by the critical coupling condition and primarily reflects the photonic Hopfield coefficient (see Sec.~\ref{subsec:Hopfield}). As the polariton becomes increasingly photonic, the absorption decreases rapidly. Conversely, in the more hybridized or slightly excitonic regions, the absorption remains high, as shown in Fig.~\ref{fig:S6}(b). This is consistent with the larger photonic linewidth, $\kappa=4.5\,\mathrm{meV}$, compared to the excitonic linewidth, $\Gamma=1\,\mathrm{meV}$, used in this study. In the nonlinear regime, the absorption profile change, with a splitting between the branches and an apparent abrupt cutoff along P$_2$ (see Fig.~\ref{fig:S6}(c)). While initially surprising, it is consistent with the behavior of the photonic Hopfield coefficient, as shown in Fig.~\ref{fig:S6}(d). The key difference from the low-density regime is the shift in the polariton energies, while the Hopfield coefficients remain essentially unchanged. This results in a splitting in absorption between P$_1$ and P$_2$ that was absent at low density.

\subsection{Tuning the electron-photon coupling strength}
\begin{figure}[!t]
     \centering
     \includegraphics[width=1\textwidth]{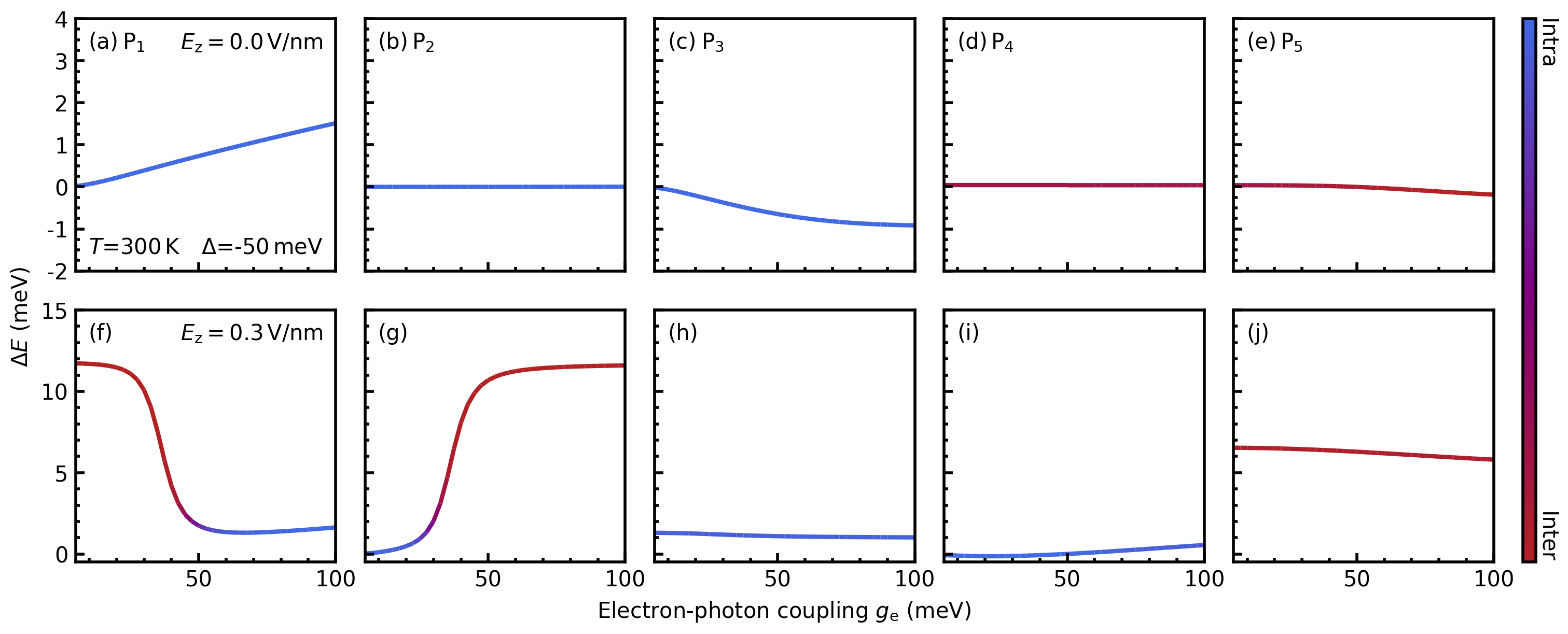}
     \caption{(a)-(e) Energy shifts of the respective lowest five polariton branches of 2H-stacked MoS$_2$ homobilayer as a function of the electron photon coupling strength, $g_\text{e}$, for a detuning of $\Delta=-50\,\mathrm{meV}$ and at room temperature. The gradient denotes the intra/inter composition of just the excitonic part of the polariton branches. (f)-(i) The same quantities as (a)-(d) including an out-of-plane electric field $E_\text{z}=0.3\,\mathrm{V/nm}$.}
     \label{fig:S8}
\end{figure}
In Fig.~5 in the main manuscript, we investigated the dependence of the magnitude of the electron-photon coupling strength on the interaction-induced energy shifts in a monolayer MoS$_2$. Here, we instead consider the corresponding homobilayer. First, we examine the polariton energy landscape as a function of $g_\text{e}$ for a red-detuned cavity at zero field and focus on the lowest five polariton branches, as presented in Fig.~\ref{fig:S7}(a). Every hybrid exciton has a two-fold degeneracy at zero field, resulting in every even polariton branch being flat and dark (P$_2$ and P$_4$), independent of the coupling strength \cite{konig2025magneto}. For increasing $g_\text{e}$, the lowest lying polariton P$_1$ shifts down in energy, initially quadratically, then linearly. It therefore behaves similar to the LP in a one-exciton-one-photon system, described by Eq.~(\ref{eq:level}). P$_3$ shifts up similar to the UP would for the monolayer. However, it moves closer to another higher-lying hybrid exciton and therefore forms an avoided crossing with P$_5$ at about $g_\text{e}=90\,\mathrm{meV}$.
When applying an out-of-plane electric field, the degeneracy of hybrid excitons is lifted. As shown in Fig.~\ref{fig:S7}(b) for $E_z=0.3$ V/nm, P$_5$ remains strongly detuned from the photon, and no nearby polariton branch is close enough to form an avoided crossing. As a result, it leads to an almost flat branch. The other polariton branches follow the expected behavior of level repulsion: as the coupling strength increases, the branches shift apart near resonance, maintaining the characteristic anti-crossing pattern, where the size of the crossing depends on the intralayer character of the underling hybrid exciton. In particular, at this field strength the lowest lying hybrid exciton is very interlayer-like, leading to the small avoided crossing between P$_1$ and P$_2$ at $g_\text{e}=35\,\mathrm{meV}$.

For the interaction-induced energy shifts, we find that saturation effects generally dominate for intralayer-like hybrid exciton polaritons due to the larger oscillator strength, while for interlayer-like polaritons the dipolar interactions are crucial due to the out-of-plane dipole moment. In the absence of an electric field, the dipolar interactions are canceled out due to the degeneracy of the anti-aligned interlayer excitons. This means that in this regime, only saturation contributes to the interaction-induced energy shift. In Figs.~\ref{fig:S8}(a)-(e) we show the energy shifts of the polariton branches P$_1$ to P$_5$. As P$_1$ behaves very similar to the LP in the monolayer case, the blueshift at room temperature shows a linear increase with coupling strength, similar to the blue line in Fig.~5(b) in the main text. P$_2$ and P$_4$ are dark, flat dark middle branches with zero photonic character and can therefore not exhibit any saturation effects, as shown by Eq.~\ref{eq:shift_psf_bilayer}. P$_3$ behaves like the UP for small coupling strengths and therefore shows a linear increasing redshift, similar to the orange line in Fig. 5(b) of the main text. As the photonic Hopfield coefficient of P$_3$ is transferred to P$_5$ due to the avoided crossing, the shift saturates. On the other hand, P$_5$ is too interlayer-like to feel any saturation shift and also remains energetically separated from the occupied states lower in energy. For higher fields, the degeneracy is lifted, enabling non-zero dipolar interactions. Figures~\ref{fig:S8}(f)–(j) show the energy shifts of the polariton branches now at a field strength of 0.3V/nm. We find that the interlayer-like polaritons experience much larger dipolar energy shifts, whereas intralayer-like polaritons are still dominated by saturation effects. Despite its large interlayer character, P$_5$ still shows only a relatively small shift, due to being too far from occupied regions.
 \bibliography{ref}